\documentclass[aps,prb,twocolumn,superscriptaddress,floatfix,showpacs,10pt,letterpaper]{revtex4}
\usepackage{graphicx, bm, amsmath, amsfonts,amssymb}

\def\c#1{\mathcal{#1}}
\def\t#1{\widetilde{#1}}

\def\al{\alpha}
\def\th{\theta}

\def\dd{\mathrm{d}}
\def\'{\prime}

\begin{document}

\title{Efficient simulation of Grassmann Tensor Product States}
\author{Zheng-Cheng Gu}
\affiliation{Kavli Institute for Theoretical Physics, University of
California, Santa Barbara, California 93106, USA }
\date{{\small \today}}

\begin{abstract}
Recently, the Grassmann-tensor-entanglement renormalization
group(GTERG) algorithm has been proposed as a generic variational approach
to study strongly correlated boson/fermion systems[Gu \emph{et al.}, arXiv:condmat/1004.2563].
However, the weakness of such a simple variational approach is that
generic Grassmann tensor product states(GTPS) with large inner
dimension $D$ will contain a large number of variational parameters
which are hard to be determined through usual minimization procedures.
In this paper, we first introduce a standard form of GTPS which
significantly simplifies the representations. Then we describe a
simple imaginary-time-evolution algorithm to efficiently update the
GTPS based on the fermion coherent state representation and show that all
the algorithms developed for usual tensor product states(TPS) can be
implemented for GTPS in a similar way. Finally, we study the
environment effect for the GTERG approach and propose a simple
method to further improve its accuracy. We demonstrate our
algorithms by studying some simple two dimensional free and interacting fermion systems on honeycomb
lattice, including both off-critical and critical cases.
\end{abstract}
\maketitle

\section{Introduction}
Since the discovery of the fractional quantum hall effect(FQHE) and high
$T_c$ cuprates, it has been realized that a large class of phases
and phase transitions can not be described by Landau symmetry
breaking theory. Enormous efforts have been made to understand the
underlying physics of these new systems during the last two decades.
It is believed that the strongly correlated nature plays an
essential role for these new phases of quantum matter. The most
successful and powerful approach to study strongly correlated systems
is to construct new classes of variational wave functions. For example, the
famous Laughlin wave function\cite{Laughlinwf} successfully explains
the quantized nature of the Hall conductance at rational filling factors. Such a new state is very different from a symmetry
breaking state and it describes a new class of order of quantum matter--
the topological order\cite{WNtop}. Although the essential
physics of high-$T_c$ is still controversial, it is believed that
the relevant low energy physics is dominated by a class of
metastable states--the resonating-valence-bond(RVB)
states\cite{AndersonRVB,PatrickRVB}. A quantitative description for
the RVB states is based on the projective wave function approach,
which was first proposed to satisfy the no-double-occupancy
constraint for the repulsive Hubbard model in the strong coupling
limit\cite{projectivewf,PatrickRVB}. It has been shown that this new class of
states can describe new phases of matter with topological order
or quantum order\cite{XGSLwf}. Later, those projective functions are
widely used to study the novel phenomena in strongly correlated
systems, including frustrated magnets\cite{YingSL} and the
fractional quantum hall states as well\cite{XGFQHEwf}.

Despite the success of projective states, they are especially
designed to describe states with particular topological order or
quantum order and it is very difficult to study the competing effect
among different orders. Therefore, it is very important to establish a
unified framework to encode different orders of quantum matter. In
Ref.\cite{GuGTPS}, a natural generalization of the projective
states, the Grassmann tensor product states(GTPS) has been proposed
as generic variational wave functions to study interacting
boson/fermion systems.

However, only local GTPS(GTPS with short range bonds)can be
efficiently simulated in an approximated\cite{GuGTPS} way. Thus, it is
also very important to understand what kind of states can be
faithfully represented in a local way. For spin/bosonic systems, Refs.
\cite{GustringTPS,VidalstringTPS} have shown that the ground states
of non-chiral topological phases, the so called string-net
condensates\cite{Levinstring}, admit a local tensor product
states(TPS) representation. Recenlty, the fermionic version of
string-net states which can describe non-chiral
topological orders in interacting fermion systems(e.g.,
fractional topological insulators) were proposed in Ref.\cite{Gufstring}. Similar to
the bosonic string-net states, the ground states of fermionic string-net models can also be
faithfully represented as GTPS since the parent Hamiltonians for
these new classes of states are described by summations of
(fermionic)commuting projectors. Moreover, it has been shown that
even for systems with chiral topological orders, the ground state wave functions admit an approximate local GTPS representation\cite{CooperGTPS,chiral1,chiral2}.
Therefore, to the best of our knowledge, the GTPS
variational approach can in principle describe all kinds of gapped
local boson/fermion systems in $1+2$D. Clearly, the advantage of
this new variational approach is that it provides a unified
description for different orders of quantum matter and allows
us to study the competing effect among different orders.

On the other hand, from the quantum information and computation
perspective, it has been shown that ground states of gapped local
Hamiltonians obey area laws. For local boson/spin systems
with translational invariance, states that satisfy such a property can be
efficiently represented by the class of so-called matrix product
states(MPS) in one dimension and by tensor product states(TPS) or
projected entangled pair states (PEPS) in higher
dimensions\cite{reviewMPS,CiracTPS}. Recently, a fermionic
generalization of those states -- the fermionic projected entangled
pair states (fPEPS) were proposed and have been bench marked in
many interesting free/interacting fermion
systems\cite{FrankfPEPS,fermionicTPS,ifPEPS,finitefPEPS,GPEPS,fPEPStJstripe,fPEPSttJ}.
In Ref.\cite{GuGTPS}, it has been shown that all fPEPS can be represented
as (local)GTPS.

Although GTPS variational ansatz is conceptually useful, the implementation in
generic strongly correlated boson/fermion systems is still not easy
since the tensor contraction for generic GTPS is an exponentially hard
problem. Similar difficulties may occur for PEPS(fPEPS)\cite{TPShard}
and many efforts have been made based on the MPS
algorithm\cite{TPSalgorithm,iPEPS,cornerMPS}. However, it is still a very big cost to handle a large system with
periodical boundary condition(PBC)\cite{TPSalgorithm}.
Alternatively, based on the concept of
renormalization\cite{LevinTRG}, the so-called tensor-entanglement
renormalization group(GTERG)\cite{GuTERG,XiangTRG1} method and its
recent developments\cite{XiangTRG2,XieTRG,LingMCTRG} are
very successful for systems with PBC. Similar to TERG, based on the
renormalization principle for Grassmann variables, the
Grassmann-tensor-entanglement renormalization group(GTERG) was
proposed in Ref.\cite{GuGTPS} to simulate physical measurements for
GTPS approximately. Nevertheless, a naive minimization procedure for
generic GTPS variational approach will still be very hard due to the
large number of variational parameters when inner dimension $D$
increases(scale as $D^3$ on honeycomb lattice and as $D^4$ on square
lattice). For TPS, it is well known that the imaginary time evolution
algorithm is the best method to solve such a problem. Hence, it is
natural to generalize the algorithm for GTPS, which is the main focus of this paper.

The rest of the paper is organized as follows:
In Section II, we present a standard form of GTPS, which only
contains one species of Grassmann variable for each inner index and
significantly simplifies the representation for numerical
calculations. In Section III, we first give a brief review about the
concept of the imaginary time evolution algorithm for TPS and then
present the detailed implementation for GTPS. Finally, we demonstrate the algorithm for a
simple spinless fermion system on honeycomb lattice, including both
off-critical and critical cases. In addition, we study a spinless fermion system with attractive
interactions on honeycomb lattice and predict a $p+ip$ superconducting ground state. We
benchmark the ground state energy with exact diagonalization calculation and find a very good agreement.
In Section IV, we describe the
environment effect of the GTERG algorithm and present a simple
improved algorithm. We implement the algorithm to a critical free
fermion system on honeycomb lattice and find a significant
improvement.
Finally, we briefly summarize our results and discuss
possible future developments along this direction.

\section{Standard form for GTPS}
%Since most interesting model Hamiltonians for interacting fermion
%systems, such as Hubbard model, $t$-$J$ model have translational
%invariance, we will limit our discussion for translational invariant
%systems throughout the whole paper, although GTPS can be generically
%defined for systems without translational invariance.

In this section, we will introduce a standard form to represent
GTPS. In the standard form, each link only associates with one
Grasmann variable, thus, the representation in the numerical
calculations will be simplified significantly.

\begin{figure}
\begin{center}
\includegraphics[scale=0.35]{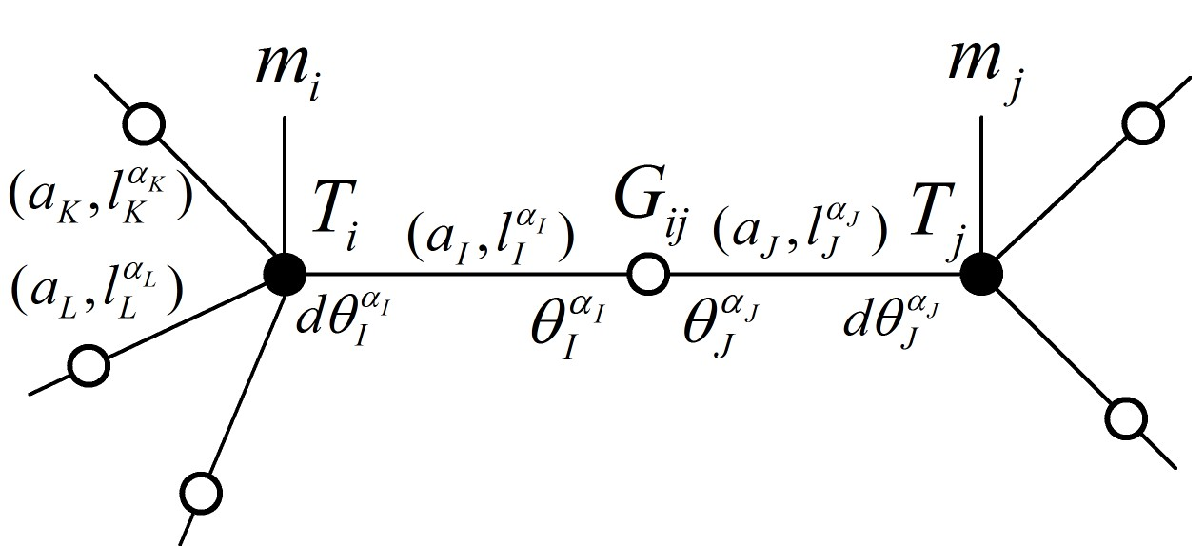}
\end{center}
\caption{ \label{GrassmannTG}A graphic representation for the GTPS.
The open circle which connects to links $I$ and $J$ represents the
Grassmann metric $G_{ij,a_Ia_J}$. The solid circle on the physical
site $i$ represents the Grassmann tensors $T^{m_i}_{i;a_Ka_L...}$.
The Grassmann numbers $\th^\al_{I}$ associate with the Grasmann
metric $G_{ij,a_Ia_J}$ and the dual Grassmann numbers associate with
the Grassmann tensor $T^{m_i}_{i;a_Ka_L...}$. There are a pair of
indices $(a_I,\{l_I^{\alpha_I}\})$ live on link $I$. $a_I$ is called
the bosonic index while $\{ l_I^{\alpha_I} \}=0,1$ are called the
fermionic indices.}
\end{figure}

Let us recall the generic GTPS wavefunctions(defined in the usual
Fork basis):
\begin{align}
\Psi(\{m_i\})= \sum_{\{a_I\}} P_0\int \prod_i T^{m_i}_{i;a_Ka_L...}
\prod_{ij} G_{ij;a_I a_J} , \label{GTPS}
\end{align}
where
\begin{align}
&  T_{i;a_Ka_L...}^{m_i}= \sum_{\{l^{\al_K}_K\} \{l^{\al_L}_L\}...}
{\rm{T}}^{m_i;\{l^{\al_K}_K\}\{l^{\al_L}_L\}...}_{i;a_Ka_L...}\prod_{I\in
i} \widetilde{\prod_{\al_I}} (\dd \th^{\al_I}_{I})^{l^{\al_I}_I}
\nonumber\\
&  G_{ij;a_I a_J}=\sum_{\{l^{\al_I}_I\}\{l^{\al_J}_J\}}
{\rm{G}}^{\{l^{\al_I}_I\}\{l^{\al_J}_J\}}_{ij;a_I a_J} \prod_{\al_J}
(\th^{\al_J}_{J})^{l^{\al_J}_J} \prod_{\al_I}
(\th^{\al_I}_{I})^{l^{\al_I}_I}\label{TG}
\end{align}
Here $i,j,\cdots$ label different physical sites, $I,J,\cdots$ label
different links and $I \in i$ means the link $I$ connects to the
site $i$(See in Fig. \ref{GrassmannTG}, any link $I$ uniquely
belongs to one physical site $i$). On each link $I$, $a_{I}$ labels
the bosonic inner indices, $l_{I}^{\al_I}=0,1$ labels the fermionic
inner indices and $\al_{I}$ labels different species of Grassmann
variables. $m_i$ is the physical index. $\th^{\al_I}_{I}$ and $\dd
\th^{\al_I}_{I}$ are Grassmann numbers and dual Grassmann numbers
that satisfy the standard Grassmann algebra:
\begin{align}
 \th^{\al_I}_{I} \th^{\al^\prime_{I^\prime}}_{I^\prime}&=-\th^{\al^\prime_{I^\prime}}_{I^\prime}\th^{\al_I}_{I},
& \dd \th^{\al_I}_{I}\dd
\th^{\al^\prime_{I^\prime}}_{I^\prime}&=-\dd
\th^{\al^\prime_{I^\prime}}_{I^\prime} \dd \th^{\al_I}_{I},
\nonumber\\
\int \dd \th^{\al_I}_{I} \th^{\al^\prime_{I^\prime}}_{I^\prime}
&=\delta_{I,I^\prime}\delta_{\al_I \al^\prime_{I^\prime}} & \int \dd
\th^{\al_I}_{I} 1&=0 .
\end{align}
 Note
that $\prod_i$ and $\widetilde {\prod_i}$ have opposite orders:
\begin{align}
 \prod_i \th_i \equiv \th_1 \th_2 \th_3...\ \ \ \ \ \
 \widetilde {\prod_i}  \th_i \equiv ...  \th_3 \th_2 \th_1  .
\end{align}
The symbol $P_0$ represents a projection of the result of the
integral to the term containing no Grassmann variables
$\th^{\al_I}_{I}$.

For fermion(electron) systems, the physical index $m_i$ in a local
Hilbert space is always associated with a definite fermion parity
$P_f(m_i)=\pm 1$. Hence, we can impose the following constraints to
issue that Eq. (\ref{GTPS}) does represent fermion wavefunctions.
\begin{align}
& \sum_{I\in i}\sum_{\al_I} l^{\al_I}_I =\text{odd, if
$P_f(m_i)=-1$} \nonumber\\
& \sum_{I\in i}\sum_{\al_I} l^{\al_I}_I =\text{even,if
$P_f(m_i)=1$} \nonumber \\
& \sum_{\al_I} l^{\al_I}_I+ \sum_{\al_J} l^{\al_J}_J = \text{even}
\label{constraint}
\end{align}

Although the original form of Eq.(\ref{GTPS}) provides us a good
physical insight of the state, especially for strongly correlated
systems from projective constructions, it is not an efficient
representation for numerical simulations. In the following we will
derive the standard form of GTPS to simplify the representation.

\begin{figure}
\begin{center}
\includegraphics[scale=0.42]{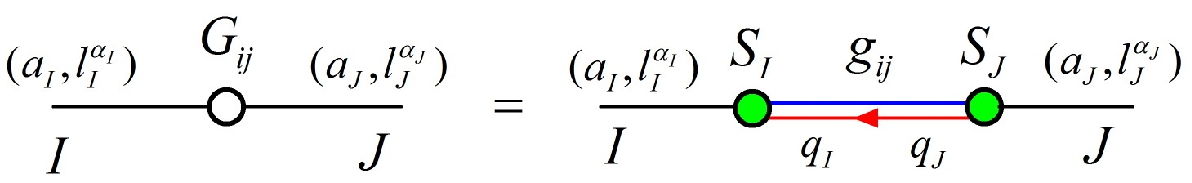}
\end{center}
\caption{ \label{tensor} A graphic representation of the
decomposition Eq. (\ref{decompose}). We use double lines(red and
blue) to represent the standard metric $g_{ij}$, which is the $Z_2$
graded version of the canonical delta function. The blue line
represents the channel with no inner fermion($n_I=n_J=0$) and the
red line represents the channel with one inner fermion($n_I=n_J=1$)
for the standard metric. The arrow of the red line represents the
ordering for the dual Grassmann variables $\dd\theta_I$ and
$\dd\theta_J$. We notice that the standard metric $g_{ij}$ only has one
species of Grassmann variable despite the original Grassmann metric
$G_{ij}$ contains many species of Grassmann variables on its link
$I$ and $J$(labeled by $\alpha_I$ and $\alpha_J$). Here $q_I,q_J$
and $a_I,a_J$ are bosonic indices.}
\end{figure}

By using the Grassmann version of the singular-value-decomposition(GSVD)
method proposed in Ref.\cite{GuGTPS}, under the constraint
$\sum_{\al_I} l^{\al_I}_I+ \sum_{\al_J} l^{\al_J}_J = \text{even}$,
we can decompose $G_{ij;a_Ia_J}$ into(see Fig. \ref{tensor}):
\begin{align}
G_{ij;a_Ia_J}=\sum_{p_I p_J}P_0^\prime\int g_{ij;q_Iq_J}
S_{I;a_Iq_I} S_{J;a_Jq_J}\label{decompose}
\end{align}
with
\begin{align}
g_{ij;q_Iq_J}&=(1+\dd \th_{J}\dd \th_{I})\delta_{q_Iq_J}\nonumber\\
S_{I;a_Iq_I}&=\sum_{\{l_I^{\al_I}\}n_I}{\rm{S}}_{I;a_Iq_I}^{\{l_I^{\al_I}\}n_I}
\left[\prod_{\al_I} (\th^{\al_I}_{I})^{l^{\al_I}_I}\right]\th_I^{n_I}\nonumber\\
S_{J;a_Jq_J}&=\sum_{\{l_J^{\al_J}\}n_J}{\rm{S}}_{J;a_Jq_J}^{\{l_J^{\al_J}\}n_J}
\left[\prod_{\al_J}
(\th^{\al_J}_{J})^{l^{\al_J}_J}\right]\th_J^{n_J}
\end{align}
and
\begin{align}
n_I =n_J =\sum_{\al_I} l^{\al_I}_I   \text{ mod } 2  = \sum_{\al_J}
l^{\al_J}_J   \text{ mod } 2,
\end{align}
where
${\rm{S}}_{I;a_Iq_I}^{\{l_I^{\al_I}\}n_I}=\sqrt{\Lambda_{q_I,n_I}}U_{a_I{\{l_I^{\al_I}\}};q_I,n_I}$
and
${\rm{S}}_{J;a_Jq_J}^{\{l_J^{\al_J}\}n_J}=\sqrt{\Lambda_{q_J,n_J}}V_{a_J{\{l_J^{\al_J}\}};q_J,n_J}$
are determined by the singular-value-decomposition(SVD) for the
matrix
$M_{a_I{\{l_I^{\al_I}\};a_J{\{l_J^{\al_J}\}}}}={\rm{G}}^{\{l^{\al_I}_I\}\{l^{\al_J}_J\}}_{ij;a_I
a_J}$ with $ M=U \Lambda V^T$. (Notice that the constraint
$\sum_{\al_I} l^{\al_I}_I+ \sum_{\al_J} l^{\al_J}_J = \text{even}$
implies a $Z_2$ symmetry for the matrix $M$ which can be block
diagonalized, with each sector labeled as $n_I=n_J=0$ or $1$.)
Again, the symbol $P_0^\prime$ represents a projection of the result
of the integral to the term containing no Grassmann number $\th_I$.
We call $g_{ij;q_Iq_J}$ the standard metric for GTPS, which is the
Grassmann generalization of the canonical delta function
$\delta_{q_Iq_J}$.

Put Eq.(\ref{decompose}) into Eq.(\ref{GTPS}), we have:
\begin{align}
\Psi(\{m_i\})&= \sum_{\{a_I\},\{q_I\}}P_0P_0^\prime \int  \prod_{ij}
g_{ij;q_I q_J}\nonumber\\&\times\prod_i T^{m_i}_{i;a_Ka_L...}
\prod_{I} S_{I;a_Iq_I} , \label{newGTPS}
\end{align}
The Grassmann matrices $S_{I;a_Iq_I}$ defined on all links contain
even number of Grassmann numbers and they commute with each other.
Such a property allows us to regroup them as:
\begin{align}
\prod_{I} S_{I;a_Iq_I}=\widetilde{\prod_{i}}\widetilde{\prod_{I\in
i}} S_{I;a_Iq_I}, \label{regroup}
\end{align}
To derive the above expression, we use the fact that each link $I$ uniquely belongs to a site $i$. $\widetilde{\prod}$ defined
here has opposite orders according to that defined in Eq.
(\ref{TG})

\begin{figure}
\begin{center}
\includegraphics[scale=0.4]{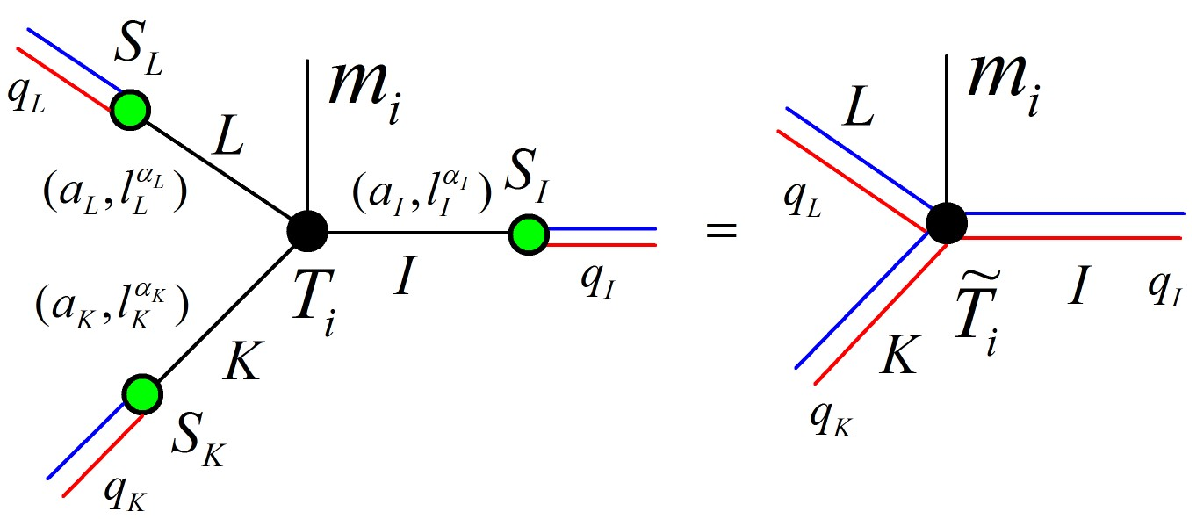}
\end{center}
\caption{ \label{newT} A graphic representation for the standard
Grassmann tensor $\t T_i$, which is a combination of the Grassmann
tensor $T_i$ and those $S_{I(K,L)}$(green solid circles) surrounding
it.}
\end{figure}

Thus, we can integral out all the Grassmann numbers $\prod_{\al_I}
(\th^{\al_I}_{I})^{l^{\al_I}_I}$ and sum over all the bosonic
indices $\{a_I\}$ to derive a simplified wave function:
\begin{align}
\Psi(\{m_i\})= \sum_{\{q_I\}} P_0^\prime\int  \prod_{ij} g_{ij;q_I
q_J}\prod_i \t T^{m_i}_{i;q_Kq_L...}   , \label{simpleGTPS}
\end{align}
where the new Grassmann tensor $\t T_{i;a_Ka_L...}^{m_i}$ associated
with physical site $i$ can be expressed as(see Fig. \ref{newT}):
%\begin{align}
%&\t T_{i;a_Ka_L...}^{m_i} = \sum_{a_Ka_L...}\sum_{\{l^{\al_K}_K\}
%\{l^{\al_L}_L\}...}P_0\int
%{\rm{T}}^{m_i;\{l^{\al_K}_K\}\{l^{\al_L}_L\}...}_{i;a_Ka_L...}\nonumber\\
%&\times\prod_{I\in i} \widetilde{\prod_{\al_I}} (\dd
%\th^{\al_I}_{I})^{l^{\al_I}_I}\widetilde{\prod_{I^\prime\in
%i}}\sum_{\{l_{I^\prime}^{\al_{I^\prime}^\prime}\}n_{I^\prime}}{\rm{S}}_{I^\prime;a_{I^\prime}q_{I^\prime}}
%^{\{l_{I^\prime}^{\al_{I^\prime}^\prime}\}n_{I^\prime}}\left[\prod_{\al_{I^\prime}^\prime}
%(\th^{\al_{I^\prime}^\prime}_{I^\prime})^{l^{\al_{I^\prime}^\prime}_{I^\prime}}\right]\th_{I^\prime}^{n_{I^\prime}}
%\end{align}
\begin{align}
\t T_{i;q_Kq_L...}^{m_i} = \sum_{n_Kn_L...} \t
{\rm{T}}^{m_i;n_Kn_L...}_{i;q_Kq_L...}  \prod_{I\in i} \th_I^{n_I}
\end{align}
with
\begin{align}
\t
{\rm{T}}^{m_i;n_Kn_L...}_{i;q_Kq_L...}&=\sum_{a_Ka_L...}\sum_{\{l^{\al_K}_K\}
\{l^{\al_L}_L\}...}\nonumber\\
&{\rm{T}}^{m_i;\{l^{\al_K}_K\}\{l^{\al_L}_L\}...}_{i;a_Ka_L...}\prod_{I\in
i} {\rm{S}}_{I;a_{I}q_{I}} ^{\{l_{I}^{\al_{I}}\}n_{I}}
\end{align}

\begin{figure}
\begin{center}
\includegraphics[scale=0.35]{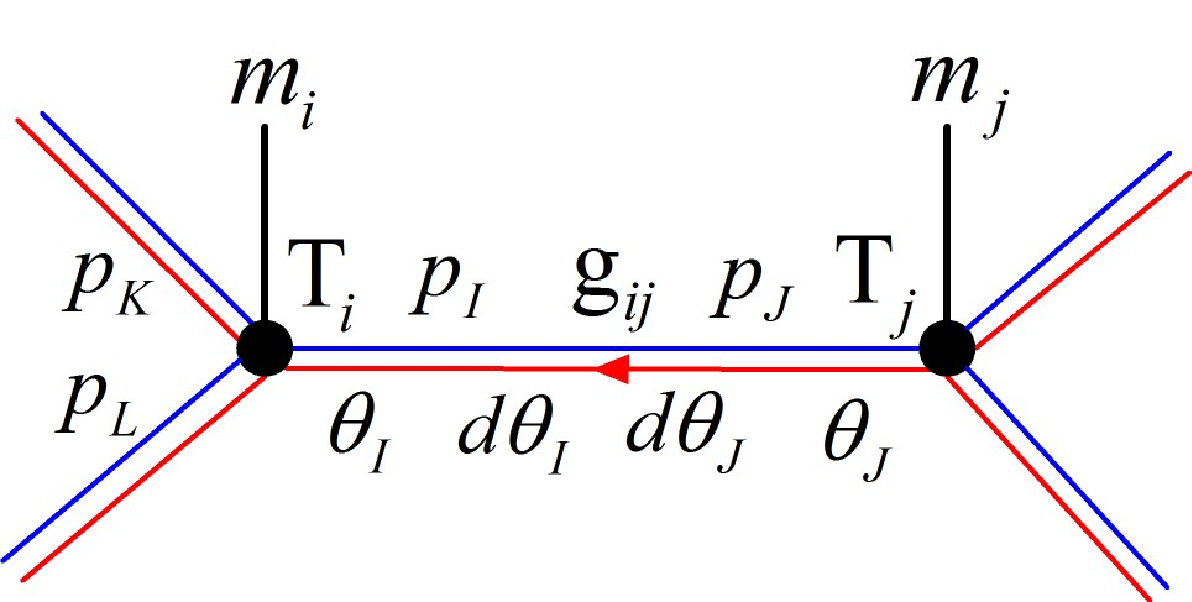}
\end{center}
\caption{ \label{simpleTG} A graphic representation for the standard
form of GTPS.  The solid circle on the physical site $i$ represents
the standard Grassmann tensors $\mathbf{T}^{m_i}_{i;p_Kp_L...}$. The
link index $p_I$ has a definite fermion parity $P_f$ and we use the
blue(red) line to represent the fermion parity even(odd)
$P_f(p_I)=1(-1)$ indices. If $P_f(p_I)=-1$, we associate a Grassmann
number $\th_I$ with the standard Grassmann tensor
$\mathbf{T}^{m_i}_{i;p_Kp_L...}$ while associate its dual $\dd
\th_I$ with the standard metric $\mathbf{g}_{ij,p_Ip_J}$. Since the
standard metric $\mathbf{g}_{ij,p_Ip_J}$ is actually just a
Grassmann generalization of the canonical delta function
$\delta_{p_Ip_J}$, we only need to use an arrow to specify the
ordering of the two Grassmann variables $\dd \th_J$ and $\dd
\th_I$.}
\end{figure}

We call Eq.(\ref{simpleGTPS}) the standard form(see Fig. \ref{simpleTG}) of GTPS which only
contains one species of Grassmann variable on each link. We can
further simplify the expression by grouping the bosonic index $q_I$ and
fermionic index $n_I$ into one super index $p_I=(q_I,n_I)$.
\begin{align}
\Psi(\{m_i\})= \sum_{\{p_I\}} \int  \prod_{ij} \mathbf{g}_{ij;p_I
p_J}\prod_i \mathbf{T}^{m_i}_{i;p_Kp_L...} , \label{NGTPS}
\end{align}
with
\begin{align}
\mathbf{g}_{ij;p_I p_J}&=\delta_{p_Ip_J}(\dd \th_J)^{N_f(p_J)}(\dd
\th_I)^{N_f(p_I)}\nonumber\\  \mathbf{T}^{m_i}_{i;p_Kp_L...} &= \t
{\rm{T}}^{m_i;n_Kn_L...}_{i;q_Kq_L...}  \prod_{I\in i}
(\th_I)^{N_f(p_I)},
\end{align}
where $N_f(p_I)=n_I$. Notice that the super index $p_I$ has a definite
fermion parity $P_f(p_I)=\pm 1$ and the corresponding fermion number
is determined as $N_f=\frac{P_f+1}{2}$.

%Although the original form of GTPS Eq.(\ref{GTPS}) is much easier to
%represent strongly correlated wavefunctions from projective
%construction and provide us strong physical intuition on how to
%construct new variational wavefunctions, it is not convenient for
%generic numerical search of GTPS without special design.
The new form Eq.(\ref{NGTPS}) is extremely useful in general purpose
of numerical calculations. We will use this new form to explain all
the details of our algorithms.

\section{The imaginary time evolution algorithm}
In this section, we will start with a brief review of imaginary time
evolution algorithms for TPS and then generalize all those
algorithms into GTPS. Finally, we apply the algorithms to some
fermion models on honeycomb lattice.
\subsection{A review of the algorithm of TPS}
\subsubsection{generic discussion}
Let us consider the imaginary time evolution for generic TPS
$|\Psi_0\rangle$.
\begin{equation}
|\Psi_\tau \rangle=e^{-\tau H}|\Psi_0 \rangle  \label{evolution}
\end{equation}
If we don't make any approximation, the true ground state can be
achieved in the $\tau\rightarrow\infty$ limit.
\begin{equation}
|\Psi_{GS} \rangle=\lim_{\tau\rightarrow\infty}e^{-\tau H}|\Psi_0
\rangle= \lim_{N\rightarrow\infty}e^{-N\delta\tau H}|\Psi_0
\rangle,\label{GS}
\end{equation}
where $N$ is the number of evolution steps and $\delta\tau=\tau/N$
is a sufficiently thin imaginary-time slice. However, without any
approximation, the inner dimension of TPS will increase
exponentially as the number of evolution steps increase. Hence, we need to
find out the best TPS approximation with fixed inner dimension.

Without loss of generality(WLOG), we use the honeycomb lattice geometry to explain the details
here and for the rest of the paper. To illustrate the key idea
of the algorithm, let us consider a simple case that the model
Hamiltonian $H$ only contains a summation of nearest neighbor
two-body terms:
\begin{equation}
H=\sum_{\langle ij\rangle}h_{ij}.
\end{equation}
Let us divide the Hamiltonian into three parts:
\begin{equation}
H=H_x+H_y+H_z; \quad H_\alpha=\sum_{i \in A}h_{i,i+\alpha}
\quad(\alpha=x,y,z),
\end{equation}
where $A$ labels the sublattices $A$ and $x,y,z$ label three
different nearest-neighbor directions. By applying the Trotter
expansion, we have:
\begin{equation}
e^{-\delta\tau H}=e^{-\delta\tau H_x}e^{-\delta\tau
H_y}e^{-\delta\tau H_z}+o(\delta\tau^2)
\end{equation}
Notice that each $H_\alpha$ only contains summation of commuting terms,
hence we can decompose them without error:
\begin{equation}
e^{-\delta\tau H_\alpha}=\prod _i e^{-\delta\tau h_{i,i+\alpha}}
\end{equation}

Let us expand $|\Psi_0\rangle$ under the physical basis:
\begin{align}
|\Psi_0\rangle &=\sum_{\{m_i,m_j\}}\sum_{\{a,a^\prime\}} \prod_{i\in
A} {T}_{i;abc}^{m_i}\prod_{j\in B}
{T}_{j;a^\prime b^\prime c^\prime}^{m_j}\nonumber\\
&\prod_{ij}\delta_{aa^\prime} |\{m_i,m_j\}\rangle,
\end{align}
Here $A,B$ denote two different sublattices in a unit cell and
$m_{i(j)}$ denote the physical indices on site $i(j)$, e.g.,
$m_{i(j)}=\uparrow,\downarrow$ for a spin $1/2$ system. The
canonical delta function $\delta_{aa^\prime}$ defined on link $ij$
can be regarded as the metric associated with tensor contraction,
which can be generalized to its Grassmann variable version for
fermion systems. After acting one evolution operator $e^{-\delta\tau
h_{ij}}$ onto the corresponding link $ij$, we can expand the new
state $e^{-\delta\tau h_{ij}}|\Psi_0\rangle$ as:
\begin{align}
e^{-\delta\tau
h_{ij}}|\Psi_0\rangle&=\sum_{\{m_i,m_j\}}\sum_{\{a,a^\prime\}}\sum_{m_i^\prime
m_j^\prime}E_{m_i^\prime m_j^\prime}^{m_im_j}
T_{i;abc}^{m_i^\prime}T_{j;a^\prime b^\prime c^\prime
}^{m_j^\prime}\nonumber\\&\times\prod_{i^\prime\neq i}
T_{i^\prime;edf}^{m_{i^\prime}}\prod_{j^\prime\neq
j}T_{j^\prime;e^\prime d^\prime
f^\prime}^{m_{j^\prime}}\prod_{ij}\delta_{aa^\prime}
|\{m_i,m_j\}\rangle
\end{align}
where $E_{m_i^\prime m_j^\prime}^{m_im_j}=\langle
m_im_j|e^{-\delta\tau h_{ij}}|m_i^\prime m_j^\prime\rangle$ is the
matrix element of the evolution operator on link $ij$. Using the SVD
decomposition, we can decompose the rank $6$ tensor
$T_{ij;m_im_jbcb^\prime c^\prime}=\sum_{a}\sum_{m_i^\prime
m_j^\prime}E_{m_i^\prime m_j^\prime}^{m_im_j}
T_{i;abc}^{m_i^\prime}T_{j;ab^\prime c^\prime }^{m_j^\prime}$ as:
\begin{align}
T_{ij;m_im_jbcb^\prime c^\prime}=\sum_{a a^\prime} {\c
T}^{m_i}_{i;abc}{\c T}^{m_j}_{j;a^\prime b^\prime c^\prime}\delta_{a
a^\prime}.
\end{align}
Here the indices $a,a^\prime$ have dimension $Dd^2$, where $D$ is
the inner dimension and $d$ is the physical dimension of the tensor
$T_{i;abc}^{m_i}$. After applying $e^{-\delta\tau H_x}$ on state
$|\Psi_0\rangle$, the tensor $T_i$ will be replaced by $\c T_i$:
\begin{align}
e^{-\delta\tau
H_x}|\Psi_0\rangle&=\sum_{\{m_i,m_j\}}\sum_{\{a,a^\prime\}}
\prod_{i\in A} {\c T}_{i;abc}^{m_i}\prod_{j\in B} {\c T}_{j;a^\prime
b^\prime c^\prime}^{m_j}\nonumber\\
&\times\prod_{ij}\delta_{aa^\prime} |\{m_i,m_j\}\rangle,
\end{align}

Notice that all $\{{\c T}_{i;abc}^{m_i}\}$ have enlarged inner dimension
$Dd^2$ instead of $D$ for their inner indices $a$(the bond along $x$
direction). Similarly, the dimension of indices $b,c$(the bonds
along $y,z$ directions) will also be enlarged to $Dd^2$ after
applying $e^{-\delta\tau H_y}$ and $e^{-\delta\tau H_z}$ on
$|\Psi_0\rangle$. Thus, it is easy to see that the inner dimension
will increase exponentially as evolution steps increase if we
don't make any truncation.

To solve the above difficulty, we need to find a new set of
$\{{T^\prime}_{i;abc}^{m_i}\}$ with fixed inner dimension $D$ that
minimizes the distance with $e^{-\delta\tau H_x}|\Psi_0\rangle$. If
we start from $|\Psi_0\rangle$ and evolve it by $e^{-\delta\tau
H_x}$ in a sufficiently thin time slice, the cost function $f$ takes
the form:
\begin{align}
&f(\{{T^\prime}_{i;abc}^{m_i}\})=|||\Psi_0^\prime\rangle
-e^{-\delta\tau H_x}|\Psi_0\rangle|| \nonumber\\&=\langle
\Psi_0^\prime|\Psi_0^\prime\rangle-\left(\langle
\Psi_0^\prime|e^{-\delta\tau H_x}|\Psi_0\rangle+h.c.\right) +const.,
\end{align}
where
\begin{align}
|\Psi_0^\prime\rangle &=\sum_{\{m_i,m_j\}}\sum_{\{a,a^\prime\}}
\prod_{i\in A} {T^\prime}_{i;abc}^{m_i}\prod_{j\in B}
{T^\prime}_{j;a^\prime b^\prime c^\prime}^{m_j}\nonumber\\
&\prod_{ij}\delta_{aa^\prime} |\{m_i,m_j\}\rangle,
\end{align}
$f$ is a multi-variable quadratic function of
$\{{T^\prime}_{i;x_iy_iz_i}^{m_i}\}$, hence we can use the sweep
method to minimize it. The advantage of the above algorithm is that
the Trotter error will not accumulate after long time evolution.
However, calculating the cost function $f$ explicitly is an
exponentially hard problem and we need further approximations at
this stage. Some possible methods have been proposed based on the
MPS algorithm\cite{TPSalgorithm}, but the calculational cost can
still be very big and the method has only been implemented with the
open boundary condition(OBC) so far.

\subsubsection{Translational invariant systems}
Nevertheless, for translational invariant TPS ansatz, it is possible
to develop an efficient method to simulate the cost function by using
the TERG method.

\begin{figure}
\begin{center}
\includegraphics[scale=0.45]{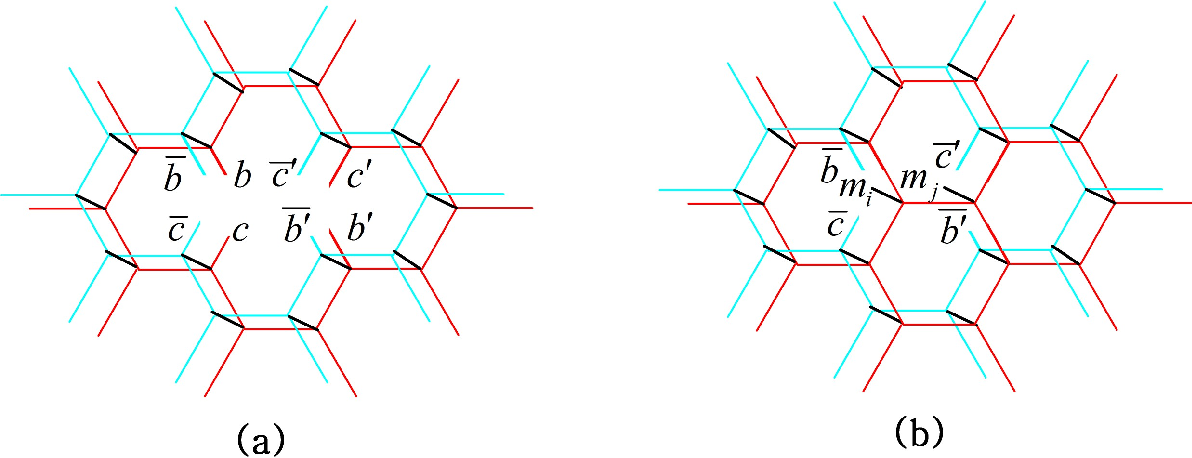}
\end{center}
\caption{A graphic representation for the environment tensor
$\rho$((a)) and $e$((b)) on honeycomb lattice.
\label{environmenttensor}}
\end{figure}

\begin{figure}
\begin{center}
\includegraphics[scale=0.25]{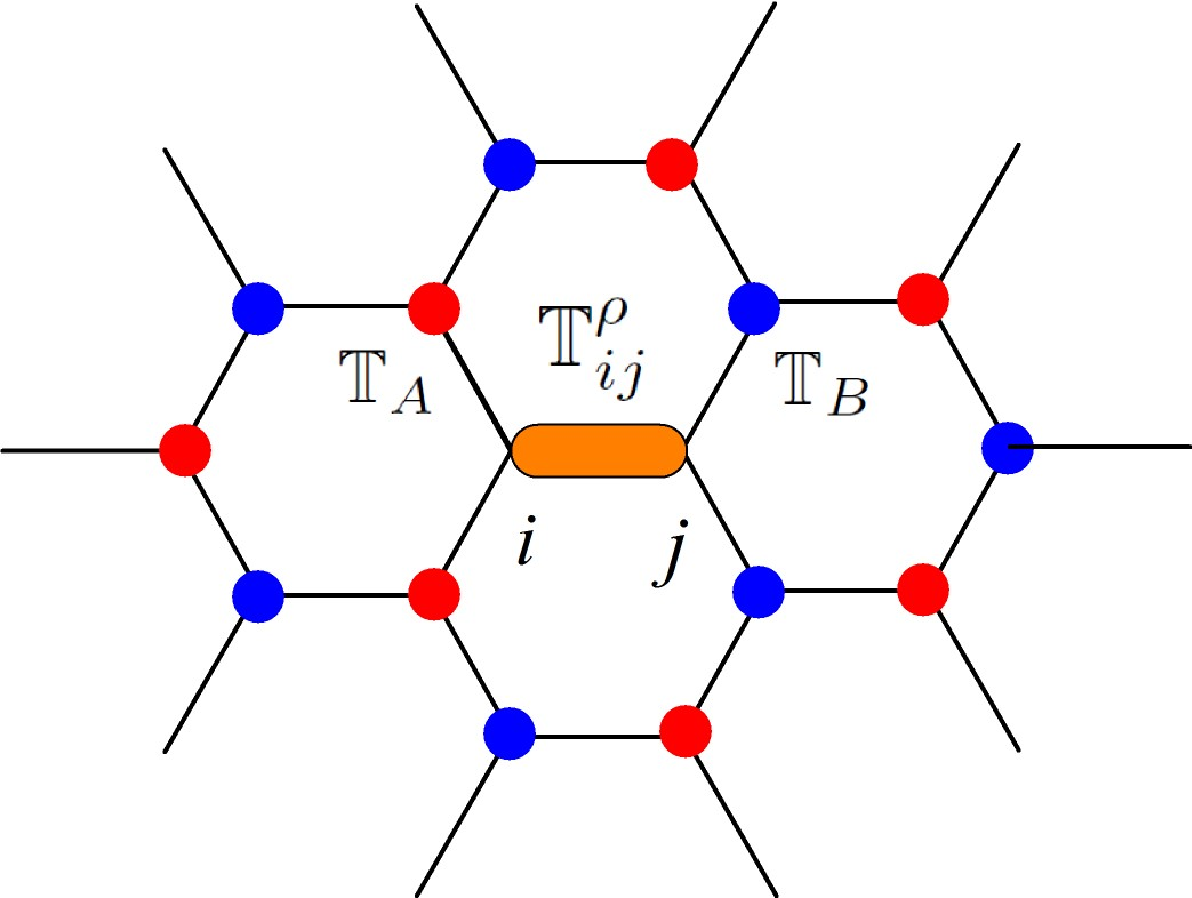}
\end{center}
\caption{A graphic representation for the tensor contraction that
allow us to compute $\rho$ on honeycomb lattice.
\label{figenvironment}}
\end{figure}

We assume that $T_i=T_A$ if $i \in A$ and $T_j=T_B$ if $j \in B$. The
cost function can be expressed as:
\begin{align}
f(T^\prime_A,T^\prime_B)&= \rho^{bcb^\prime c^\prime;\bar b \bar c
\bar b^\prime \bar c^\prime}  {\bar {T^\prime}}_{A;\bar a\bar b\bar
c}^{m_i}{T^\prime}_{A;abc}^{m_i} {\bar {T^\prime}}_{B;\bar a \bar
b^\prime\bar c^\prime}^{m_j}{T^\prime}_{B;a b^\prime
c^\prime}^{m_j}\nonumber\\& - \left(e^{\bar b \bar c\bar b^\prime
\bar c^\prime; m_im_j}{\bar {T^\prime}}_{A;\bar a\bar b\bar c}^{m_i}
{\bar {T^\prime}}_{B;\bar a\bar b^\prime\bar c^\prime}^{m_j}
+h.c.\right)\nonumber\\ &+const,\label{costf}
\end{align}
where $\rho$ and $e$(Fig. \ref{environmenttensor}) are the so called
environment tensors. Here we use the convention that all the
repeated indices will be summed over and we use $\bar T$ to represent
complex conjugate of $T$. Strictly speaking, the environment tensors
for $\rho$ and $e$ are also dependent on $T_A^\prime$ and
$T_B^\prime$, thus $f$ is no longer a quadratic multi-variable
function. However, for sufficiently thin time slice, up to
$o(\delta\tau^2)$ error(same order as Trotter error), we can replace
$T_A^\prime ,T_B^\prime$ by $T_A,T_B$ when calculating the
environment tensor $\rho$. $e$ can be derived from $\rho$:
\begin{align}
e^{\bar b \bar c\bar b^\prime \bar c^\prime; m_im_j}=
\rho^{bcb^\prime c^\prime;\bar b \bar c \bar b^\prime \bar
c^\prime}E_{m_i^\prime
m_j^\prime}^{m_im_j}{T}_{A;abc}^{m_i^\prime}{T}_{B;a b^\prime
c^\prime}^{m_j^\prime}
\end{align}
Again, repeated indices need to be summed over here. We notice that
$\rho$ can be expressed as a tensor trace of double tensors
$\mathbb{T}_{A(B)}=\sum_m \bar T_{A(B)}^m\otimes T_{A(B)}^m$ with an
impurity tensor $\mathbb{T}_{ij}$ for the link $ij$(see Fig. \ref{figenvironment}):
\begin{align}
\rho^{bcb^\prime c^\prime;\bar b \bar c \bar b^\prime \bar
c^\prime}={\rm{tTr}}[ \mathbb{T}_{ij}^\rho\otimes
\mathbb{T}_A\otimes \mathbb{T}_B \otimes \mathbb{T}_A\otimes
\mathbb{T}_B\cdots],
\end{align}
where the impurity double tensor $\mathbb{T}_{ij}$ is just a
projector:
\begin{align}
\mathbb{T}_{ij;bcb^\prime c^\prime;\bar b \bar c \bar b^\prime \bar
c^\prime}=1; \quad {\rm{others}}=0
\end{align}
Now it is easy to see that we can first decompose the impurity tensor on
the link $i,j$ to two rank $3$ impurity tensors on site $i,j$ and
then implement the usual TERG algorithm. We can also use a more
efficient but complicated way to compute $\rho$ by applying the
coarse graining procedures for all sites except sites $i,j$, as
introduced in Ref.\cite{XiangTRG2}.

\begin{figure}
\begin{center}
\includegraphics[scale=0.4]{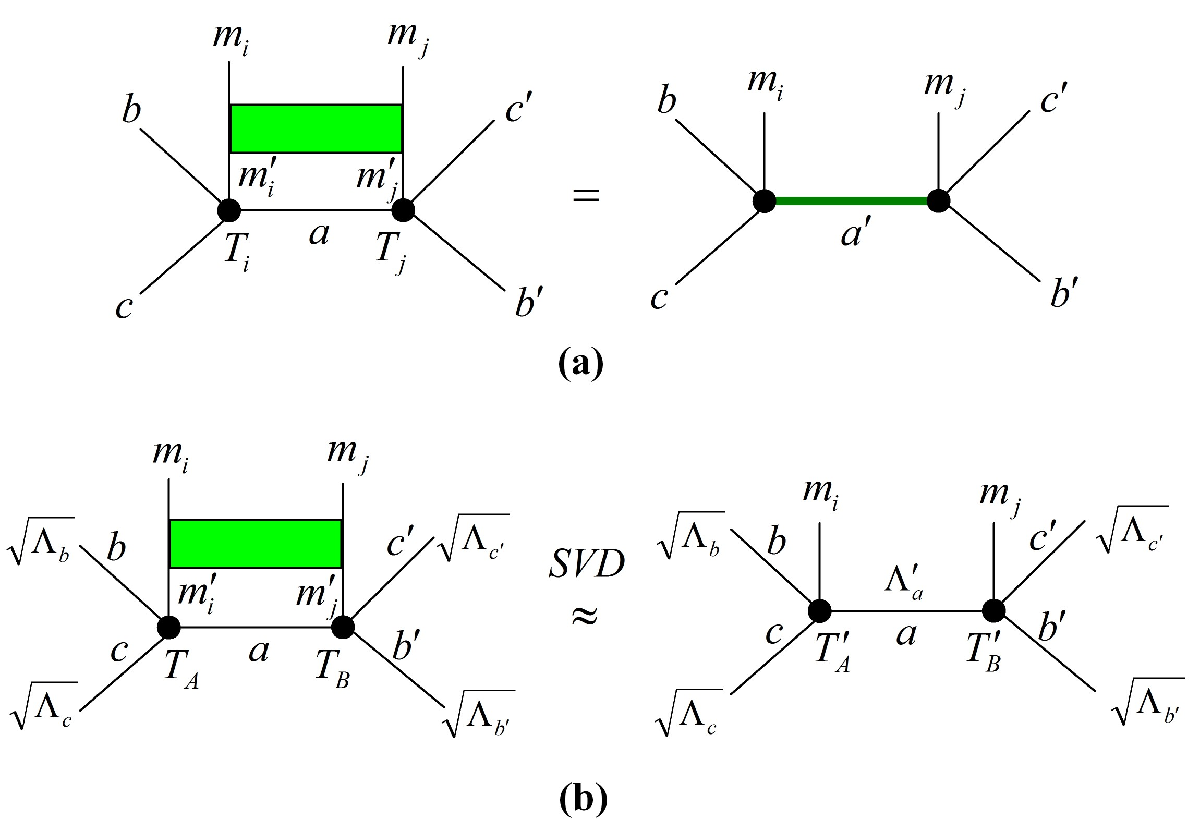}
\end{center}
\caption{(a)After applying the two body evolution operator
$e^{-\delta\tau h_{ij}}$ to the $ij$ link for TPS(with inner
dimension $D$), we get new TPS with enlarged inner dimension($Dd^2$,
$d$ is the physical dimension) for the corresponding link. (b) For
large classes of translational invariant TPS, we can reduce the
inner dimension from $Dd^2$ back to $D$ with respect to some
environment wight $\Lambda$. \label{evolution}}
\end{figure}

The above algorithm can further be simplified if we assume that the
environment tensor $\rho$ has specific forms for certain physical
systems. One interesting attempt was proposed in
Ref.\cite{XiangTRG1} by assuming that $\rho$ can be factorized as:
\begin{align}
\rho^{bcb^\prime c^\prime;\bar b \bar c \bar b^\prime \bar
c^\prime}=\Lambda_b^y
\Lambda_c^z\Lambda_{b^\prime}^y\Lambda_{c^\prime}^z  \delta_{b \bar
b}\delta_{c \bar c}\delta_{b^\prime\bar b^\prime}\delta_{c^\prime
\bar c^\prime},
\end{align}
where $\Lambda^\alpha$ is a positive weight vector defined on links
along $\alpha$ direction. The above form can always be true for
1D systems\cite{VidalMPS} due to the existence of the canonical form of
an MPS. Although the above form is not generic enough in 2D, it still
works well in many cases, especially for those systems with symmetry
breaking order. In this case, the cost function Eq.(\ref{costf}) can
be solved by SVD decomposition and keep the leading $D$th singular
values for the following matrix(see Fig. \ref{evolution}):
\begin{align}
M_{bcm_i;b^\prime c^\prime
m_j}&=\sum_{a,m_i^\prime,m_j^\prime}\sqrt{\Lambda_b^y}
\sqrt{\Lambda_c^z}\sqrt{\Lambda_{b^\prime}^y}\sqrt{\Lambda_{c^\prime}^z}
 T_{ij,m_in_jbcb^\prime
c^\prime}\nonumber\\&=\sum_{a,m_i^\prime,m_j^\prime}\sqrt{\Lambda_b^y}
\sqrt{\Lambda_c^z}\sqrt{\Lambda_{b^\prime}^y}\sqrt{\Lambda_{c^\prime}^z}\nonumber\\
&\times E_{m_i^\prime
m_j^\prime}^{m_im_j}{T}_{A;abc}^{m_i^\prime}{T}_{B;a b^\prime
c^\prime}^{m_j^\prime}\nonumber\\&\simeq \sum_{a=1}^{D}
U_{bcm_i;a}V_{b^\prime c^\prime m_j;a}\Lambda_a^\prime\label{M}
\end{align}
The new tensors $T^\prime_A$ and $T^\prime_B$ can be determined as:
\begin{align}
{T^\prime}^{m_i}_{A;abc}&=\frac{\sqrt{\Lambda_a^\prime}}{\sqrt{\Lambda_b^y}
\sqrt{\Lambda_c^z}}U_{bcm_i;a}\nonumber\\
{T^\prime}^{m_j}_{B;a^\prime b^\prime c^\prime
}&=\frac{\sqrt{\Lambda_{a^\prime}^\prime}}{\sqrt{\Lambda_{b^\prime}^y}\sqrt{\Lambda_{c^\prime}^z}}V_{b^\prime
c^\prime m_j;a^\prime}
\end{align}
Similarly as in 1D, the environment weight associated with bonds along
$x$ direction is updated as $\Lambda^x=\Lambda^\prime$. It is not
hard to understand why the above simplified algorithm works very
well for systems with symmetry breaking orders, but not in the
critical region, since in those cases the ground states are close to
product states and the entanglements between the $x,y,z$ directions in
the environment tensors become pretty weak. However, for topologically
ordered states, the environment tensors can be more complicated. For
example, the $Z_2$ topologically ordered state in the toric code model
will have an emergent $Z_2$ symmetry and can not be factorized as
a product. Hence, for topological ordered states, it is important to
calculate the full environment in the imaginary time evolution.

\subsection{Generalize to fermion(electron) systems}

In this subsection, we will show how to generalize the above
algorithms to GTPS. The key step is to introduce fermion coherent
state representation and treat the physical indices also as a
Grassmann variable. WLOG, we use a spinless fermion system as a
simple example here and for the rest part of the paper:
\begin{align}
|\eta\rangle \equiv \prod_i (1-\eta_i c_i^{\dagger}) |0\rangle,
\end{align}
where $\eta_i$ is a Grassmann number.

As already discussed in Ref.\cite{GuGTPS}, under this new
basis, the GTPS wavefunction Eq.(\ref{GTPS}) and its standard form
Eq.(\ref{NGTPS}) can be represented as:
\begin{align}
\Psi (\{ \eta_i \}) =& \sum_{\{m_i\}} \sum_{\{a_I\}} \int \prod_i
{\bar \eta_i}^{m_i} T^{m_i}_{i;a_Ka_L...} \prod_{ij} G_{ij;a_I a_J}
\nonumber\\ =&\sum_{\{m_i\}} \sum_{\{p_I\}} \int  \prod_{ij}
\mathbf{g}_{ij;p_I p_J}\prod_i {\bar
\eta_i}^{m_i}\mathbf{T}^{m_i}_{i;p_Kp_L...}
\nonumber\\=&\sum_{\{m_i\}} \sum_{\{p_I\}} \int  \prod_{ij}
\mathbf{g}_{ij;p_I p_J}\prod_i  \mathbf{\t T}^{m_i}_{i;p_Kp_L...}
\label{CGTPS}
\end{align}
where $\bar \eta_i$ is the complex conjugate of the Grassmann number
$\eta_i$ and $\mathbf{\t T}^{m_i}_{i;p_Kp_L...} = {\bar
\eta_i}^{m_i} \mathbf{T}^{m_i}_{i;p_Kp_L...}$.

On honeycomb lattice with translational invariance, we can simplify
the above expression as:
\begin{eqnarray}
&& \Psi(\{\bar \eta_i\},\{\bar \eta_j \}) \\
\nonumber &&  = \sum_{\{m_i\},\{m_j\}} \sum_{\{a\},\{a^\prime\}}
\int \prod_{\langle ij\rangle}\textbf{g}_{a a^\prime} \prod_{i\in A}
{\mathbf{\t T}}^{m_i}_{A;abc} \prod_{j\in B} {\mathbf{\t
T}}^{m_j}_{B;a^\prime b^\prime c^\prime},\label{newGTPS}
\end{eqnarray}
with
\begin{eqnarray}
{\mathbf{\t T}}^{m_{i}}_{A;abc}&=& {\rm{\t T}}^{m_{i}}_{A;abc} \bar
\eta_i^{m_i}\theta_\alpha^{N_f(a)} \theta_\beta^{N_f(b)}
\theta_\gamma^{N_f(c)},  \nonumber\\
 {\mathbf{\t T}}^{m_{j}}_{B;a^\prime b^\prime c^\prime}&=&
{\rm{\t T}}^{m_{j}}_{B;a^\prime b^\prime c^\prime} \bar
\eta_j^{m_j}\theta_{\alpha^\prime}^{N_f(a^\prime)}
\theta_{\beta^\prime}^{N_f(b^\prime)}
\theta_{\gamma^\prime}^{H_f(c^\prime)},    \nonumber\\
\textbf{g}_{aa^\prime}&=& \delta_{aa^\prime}{\dd
\theta}_\alpha^{N_f(a)} {\dd
\theta}_{\alpha^\prime}^{N_f(a^\prime)}.
\end{eqnarray}

Comparing to usual TPS, here the link indices $\{a,a^\prime\}$
always have definite fermion parity $P_f=\pm1$. $N_f=0$ when $P_f=1$
while $N_f=1$ when $P_f=-1$.

\begin{figure}
\begin{center}
\includegraphics[scale=0.4]{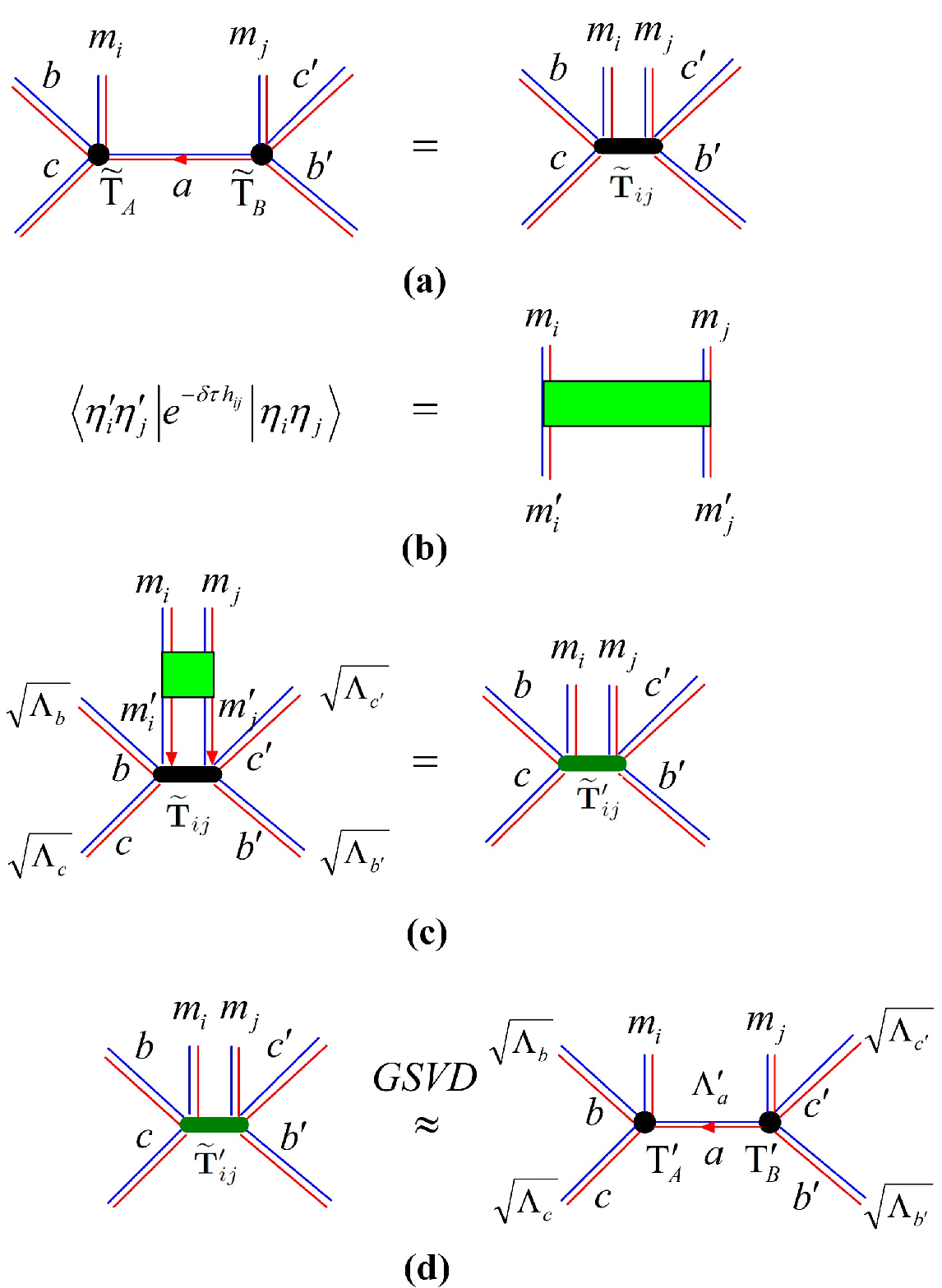}
\end{center}
\caption{A graphic representation for the simplified imaginary time
evolution algorithm for Grassmann tensor product states. In step
(a), we sum over the inner indices and integral out the Grassmann
variables on the shared link for two adjoint Grassmann tensors $
\mathbf{\t T}_A$ and $ \mathbf{\t T}_B$. In step (b), we express the
two sites evolution operator $e^{-\delta \tau h_{ij}}$ under the fermion
coherent basis. In step (c) and (d), we apply the evolution operator
$e^{-\delta \tau h_{ij}}$ to the two sites Grassmann tensor $
\mathbf{\t T}_{ij}$ and derive a new two sites Grassmann tensor $
\mathbf{\t T}_{ij}^\prime$. Then we approximate it by two new
adjoint Grassmann tensors $ \mathbf{\t T}_A^\prime$ and $ \mathbf{\t
T}_B^\prime$ with respect to the environment weights $\Lambda$. We
also use double lines to label different fermion parities for the
physical indices. For the simple spinless fermion example, the red line
represents one fermion state and the blue line represents the empty
state. The metric $(1+\eta_i\bar\eta_i)$ is the standard metric for
the physical indices. \label{fevolution}}
\end{figure}

Let us start from the simplest case with the assumption that the
environment can also be approximately represented by some weights. In this case,
the imaginary time evolution for GTPS can be reduced to an SVD
problem of Grassmann variables. The Grassmann version of the matrix
$M$ in Eq.(\ref{M}) can be constructed as followings(see Fig. \ref{fevolution}):

(a) Let us first sum over the bond indices $a,a^\prime$ and integrate
out the Grassmann variables $\th_\alpha,\th_{\alpha^\prime}$:
\begin{widetext}
\begin{align}
\mathbf{\t T}_{ij;m_im_jbcb^\prime c^\prime}&=\sum_{aa^\prime}\int
\mathbf{g}_{aa^\prime}\mathbf{\t T}_{abc}^{m_i}\mathbf{\t
T}_{a^\prime b^\prime c^\prime}^{m_j}\nonumber\\&= \sum_a\int {\dd
\theta}_\alpha^{N_f(a)} {\dd \theta}_{\alpha^\prime}^{N_f(a)}{\rm{\t
T}}^{m_{i}}_{A;abc} \bar\eta_i^{m_i}\theta_\alpha^{N_f(a)}
\theta_\beta^{N_f(b)} \theta_\gamma^{N_f(c)} {\rm{\t
T}}^{m_{j}}_{B;a b^\prime c^\prime} \bar
\eta_j^{m_j}\theta_{\alpha^\prime}^{N_f(a)}
\theta_{\beta^\prime}^{N_f(b^\prime)}
\theta_{\gamma^\prime}^{H_f(c^\prime)}\nonumber\\
&= \sum_a
(-)^{\left[m_i+m_j\right]N_f(a)}(-)^{m_j\left[m_i+N_f(b)+N_f(c)\right]}{
\rm{\t T}}^{m_{i}}_{A;abc}{\rm{\t T}}^{m_{j}}_{B;a b^\prime
c^\prime} \bar \eta_j^{m_j}\bar \eta_i^{m_i} \theta_\beta^{N_f(b)}
\theta_\gamma^{N_f(c)}
 \theta_{\beta^\prime}^{N_f(b^\prime)}
\theta_{\gamma^\prime}^{N_f(c^\prime)}.\label{Gintegral}
\end{align}
\end{widetext}
Here the sign factor
$(-)^{\left[m_i+m_j\right]N_f(a)}(-)^{m_j\left[m_i+N_f(b)+N_f(c)\right]}$
comes from the anti-commutating relations of Grassmann variables.

(b) Next we derive the matrix element of the evolution operator
under fermion coherent state representation. Let us first calculate
them under the usual Fock basis
$(c_j^\dagger)^{m_j}(c_i^\dagger)^{m_i}|0\rangle$ with
$m_i,m_j=0,1$. Let us define:
\begin{eqnarray}
E_{m_i^\prime  m_j^\prime}^{m_i  m_j}=\langle
0|(c_i)^{m_i}(c_j)^{m_j}e^{-\delta\tau h_{ij}}
(c_j^\dagger)^{m_j^\prime}(c_i^\dagger)^{m_i^\prime}|0\rangle
\end{eqnarray}
Thus, we can expand $e^{-\delta\tau h_{ij}}$ as:
\begin{align}
e^{-\delta\tau h_{ij}}=\sum_{m_im_j;m_i^\prime
m_j^\prime}E_{m_i^\prime m_j^\prime}^{m_i
m_j}(c_j^\dagger)^{m_j}(c_i^\dagger)^{m_i}|0\rangle\langle
0|(c_i)^{m_i^\prime}(c_j)^{m_j^\prime}
\end{align}
In the fermion coherent state basis, we have:
\begin{align}
&\langle \eta^\prime_i,\eta^\prime_j|e^{-\delta\tau h_{ij}}|
\eta_i,\eta_j \rangle \nonumber\\=&\sum_{m_im_j;m_i^\prime
m_j^\prime}E_{m_i^\prime m_j^\prime}^{m_i m_j}(\bar
\eta^\prime_j)^{m_j}(\bar \eta^\prime_i)^{m_i} (
\eta_i)^{m_i^\prime}( \eta_j)^{m_j^\prime}\label{corepresentation}
\end{align}

(c) Then we can evolve the state $\Psi$ to a new state
$\Psi^\prime$:
\begin{align}
\Psi^\prime(\{\bar \eta_i^\prime\},\{\bar \eta_j^\prime \}) &=\int
\dd \bar\eta_i\dd\eta_i\dd \bar\eta_j\dd\eta_j(1+\eta_i\bar\eta_i)(1+\eta_j\bar\eta_j)\nonumber\\
&\times \langle \eta^\prime_i,\eta^\prime_j|e^{-\delta\tau h_{ij}}|
\eta_i,\eta_j \rangle \Psi(\{\bar \eta_i\},\{\bar \eta_j
\}).\label{evolution}
\end{align}
Put Eq.(\ref{Gintegral}) and Eq.(\ref{corepresentation}) into the
above equation, it is easy to derive the Grassmann version of the
rank $6$ tensor $T_{ij}$ defined on the link $ij$. We have:
\begin{widetext}
\begin{align}
 \mathbf{\t T}_{ij;m_im_jbcb^\prime c^\prime}^\prime&= \sum_{m_i^\prime
 m_j^\prime a}
(-)^{\left[m_i^\prime+m_j^\prime\right]N_f(a)}(-)^{m_j^\prime\left[m_i^\prime+N_f(b)+N_f(c)\right]}
 E_{m_i^\prime m_j^\prime}^{m_i m_j}{ \rm{\t
T}}^{m_{i}^\prime}_{A;abc}{\rm{\t T}}^{m_{j}^\prime}_{B;a b^\prime
c^\prime} {(\bar\eta_j^\prime)}^{m_j} {(\bar \eta_i^\prime) }^{m_i}
\theta_\beta^{N_f(b)} \theta_\gamma^{N_f(c)}
 \theta_{\beta^\prime}^{N_f(b^\prime)}
\theta_{\gamma^\prime}^{N_f(c^\prime)}\nonumber\\
&= \sum_{m_i^\prime m_j^\prime a}
(-)^{\left[m_i^\prime+m_j^\prime\right]N_f(a)}
(-)^{m_j^\prime\left[m_i^\prime+N_f(b)+N_f(c)\right]}(-)^{m_j\left[m_i+N_f(b)+N_f(c)\right]}
 \nonumber\\ & \times E_{m_i^\prime m_j^\prime}^{m_i m_j}{ \rm{\t
T}}^{m_{i}^\prime}_{A;abc}{\rm{\t T}}^{m_{j}^\prime}_{B;a b^\prime
c^\prime} {(\bar \eta_i^\prime) }^{m_i} \theta_\beta^{N_f(b)}
\theta_\gamma^{N_f(c)} {(\bar\eta_j^\prime)}^{m_j}
\theta_{\beta^\prime}^{N_f(b^\prime)}
\theta_{\gamma^\prime}^{N_f(c^\prime)}.
\end{align}
\end{widetext}
We notice that there will be an extra sign factor
$(-)^{m_j\left[m_i+N_f(b)+N_f(c)\right]}$ after we reorder these
Grassmann variables.

(d) Finally, we can define the Grassmann generalization of the $M$
matrix after putting the environment weight for all the inner
indices.
\begin{align}
\mathbf{M}_{bcm_i;b^\prime c^\prime m_j} &= {\rm{M}}_{bcm_i;b^\prime
c^\prime m_j} {(\bar \eta_i^\prime) }^{m_i} \theta_\beta^{N_f(b)}
\theta_\gamma^{N_f(c)} \nonumber \\&\times
{(\bar\eta_j^\prime)}^{m_j} \theta_{\beta^\prime}^{N_f(b^\prime)}
\theta_{\gamma^\prime}^{N_f(c^\prime)}.
\end{align}
where the coefficient matrix $\rm{M}$ reads:
\begin{align}
{\rm{M}}_{bcm_i;b^\prime c^\prime m_j}  &= \sum_{am_i^\prime
m_j^\prime}\sqrt{\Lambda_b^y}
\sqrt{\Lambda_c^z}\sqrt{\Lambda_{b^\prime}^y}\sqrt{\Lambda_{c^\prime}^z}\nonumber\\&\times
(-)^{\left[m_i^\prime+m_j^\prime\right]N_f(a)}
(-)^{m_j^\prime\left[m_i^\prime+N_f(b)+N_f(c)\right]} \nonumber\\
&\times (-)^{m_j\left[m_i+N_f(b)+N_f(c)\right]}  E_{m_i^\prime
m_j^\prime}^{m_i m_j}{ \rm{\t T}}^{m_{i}^\prime}_{A;abc}{\rm{\t
T}}^{m_{j}^\prime}_{B;a b^\prime c^\prime}
\end{align}
Since $h_{ij}$ is a local fermionic operator, it will always contain an
even number of fermion operators. As a result, the nonzero elements
of the $\rm{M}$ matrix will always contain an even number of Grassmann
variables and we can apply GSVD as discussed before.  We keep the
largest $D$ eigenvalues:
\begin{align}
{\rm{M}}_{bcm_i;b^\prime c^\prime m_j}=\sum_a
U_{bcm_i;a}\Lambda_a^\prime V_{b^\prime c^\prime m_j;a}
\end{align}
The coefficient matrix $\rm{M}$ will have a block diagonal structure,
hence the new index $a$ will have a definite fermion parity
$P_f(a)=P_f(b)P_f(c)=P_f(b^\prime)P_f(c^\prime)$.

Similar to the usual TPS cases, the new Grassmann tensors $ \mathbf{\t
T}_i^\prime$ and $\mathbf{\t T}_j^\prime$ have the form
\begin{align}
{\mathbf{\t
{T^\prime}}}^{m_i}_{A;abc}&=\frac{\sqrt{\Lambda_a^\prime}}{\sqrt{\Lambda_b^y}
\sqrt{\Lambda_c^z}}U_{bcm_i;a} \theta_\alpha^{N_f(a)}{(\bar
\eta_i^\prime) }^{m_i} \theta_\beta^{N_f(b)}
\theta_\gamma^{N_f(c)}\nonumber\\
&={{\rm{\t T}}}^{\prime m_i}_{A;abc}{(\bar \eta_i^\prime)
}^{m_i}\theta_\alpha^{N_f(a)} \theta_\beta^{N_f(b)}
\theta_\gamma^{N_f(c)}\nonumber\\
{\mathbf{\t {T^\prime}}}^{m_j}_{B;a^\prime b^\prime c^\prime
}&=\frac{\sqrt{\Lambda_{a^\prime}^\prime}}{\sqrt{\Lambda_{b^\prime}^y}\sqrt{\Lambda_{c^\prime}^z}}V_{b^\prime
c^\prime
m_j;a^\prime}\theta_{\alpha^\prime}^{N_f(a^\prime)}{(\bar\eta_j^\prime)}^{m_j}
\theta_{\beta^\prime}^{N_f(b^\prime)}
\theta_{\gamma^\prime}^{N_f(c^\prime)}\nonumber\\ &= {{\rm{\t
T}}}^{\prime m_j}_{B;a^\prime b^\prime c^\prime
}{(\bar\eta_j^\prime)}^{m_j}\theta_{\alpha^\prime}^{N_f(a^\prime)}
\theta_{\beta^\prime}^{N_f(b^\prime)}
\theta_{\gamma^\prime}^{N_f(c^\prime)}
\end{align}

However, due to the reordering of Grassmann variables, extra signs
appear in the new tensors ${\rm{\t T}}_A^\prime$ and ${\rm{\t
T}}_B^\prime$.
\begin{align}
{{\rm{\t T}}}^{\prime
m_i}_{A;abc}&=(-)^{m_iN_f(a)}\frac{\sqrt{\Lambda_a^\prime}}{\sqrt{\Lambda_b^y}
\sqrt{\Lambda_c^z}}U_{bcm_i;a}  \nonumber\\
{{\rm{\t T}}}^{\prime m_j}_{B;a^\prime b^\prime c^\prime
}&=(-)^{m_jN_f(a^\prime)}\frac{\sqrt{\Lambda_{a^\prime}^\prime}}
{\sqrt{\Lambda_{b^\prime}^y}\sqrt{\Lambda_{c^\prime}^z}}V_{b^\prime
c^\prime m_j;a^\prime}
\end{align}
Again $\Lambda_a^\prime$ is used as the new environment weight for
all links along the $x$ direction.

The full environment tensors can be very similar as those in the usual TPS
case. The Grassmann version of the environment tensor $\rho$ can be
efficiently simulated by GTPS. The environment tensor $e$ can be
calculated from $\rho$ and $\mathbf{\t T}_{ij}^\prime$. The cost
function can be derived from Eq.(\ref{costf}) after replacing
$\rho,e$ and $T$ with their Grassmann version. After we contract the
tensor-net and integrate out all the Grassmann variables, it can be
reduced to a usual multi-variable quadratic minimization problem for
the coefficient tensors $\rm{\t T}^\prime_A$ and $\rm{\t
T}^\prime_B$, and we can solve it by using the sweep method. Although
the algorithm with full environment is general, it is still very
time consuming and a much more efficient method is very desired.
Nevertheless, it turns out that the simple updated method works very
well in many cases and we will focus on the application of this
method in this paper.

\subsection{A free fermion example}
In this subsection, we demonstrate the above algorithm by studying a
free fermion model on honeycomb lattice. We consider the following
(spinless)fermion Hamiltonian:
\begin{align}
H=-2\Delta\sum_{\langle ij \rangle}\left(c_i^\dagger
c_j^\dagger+h.c.\right)+\mu\sum_i n_i \label{freefermion}
\end{align}

\begin{figure}
\begin{center}
\includegraphics[scale=0.33]{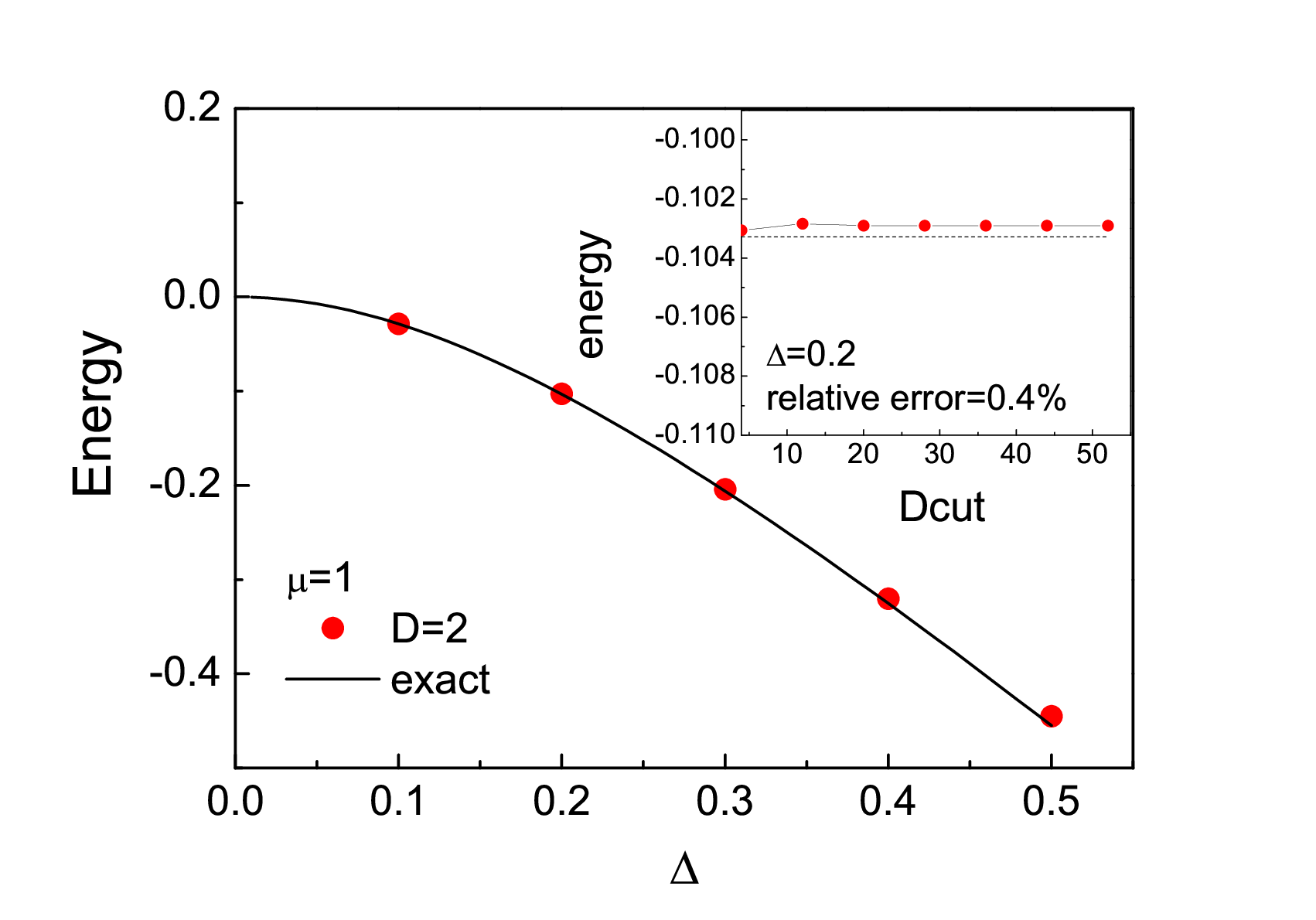}
\end{center}
\caption{(Color online)Ground state energy per site as a function of $\Delta$ for fixed
$\mu=1$. As a bench mark, we also plot the exact energy of the free
fermion Hamiltonian. (Insert:) Relative error of ground state energy
as a function of $D_{cut}$ for $\Delta=0.2$. $D_{cut}$ is the number
of singular values kept in the Grassmann SVD
decomposition\cite{CiracTPS}.\label{freegap}}
\end{figure}

We first test our algorithm in the parameter region where the system
opens a gap. For example, we can fix $\mu=1$ and take different
values for $\Delta$ from $0.1$ to $0.5$. As seen in Fig.
\ref{freegap}, even with the minimal inner dimension $D=2$,
the GTPS variational approach has already given out very good ground
state energy. Here and through out the whole paper, we fix the total
system size to be $N=2\times3^6$ sites and with PBC.(The GTEG
algorithm will allow us to reach a huge size in principle, however,
for better convergency, we choose a relatively large but not huge
size.) We note that the agreement for the ground state energy is better
for small $\Delta$, which is expected since the system becomes a
trivial vacuum state $|0\rangle$ in the limit $\Delta=0$. In the
insert of Fig. \ref{freegap}, we also plot the ground energy as
a function of $D_{cut}$ and it is shown that the energy converges for
very small $D_{cut}$(around $D_{cut}=15$, where $D_{cut}$ is the
number of singular value we keep in the GTERG algorithm). We find that
the relative error can be very small, e.g., $0.4\%$ for
$\Delta=0.2$.

\begin{figure}
\begin{center}
\includegraphics[scale=0.33]{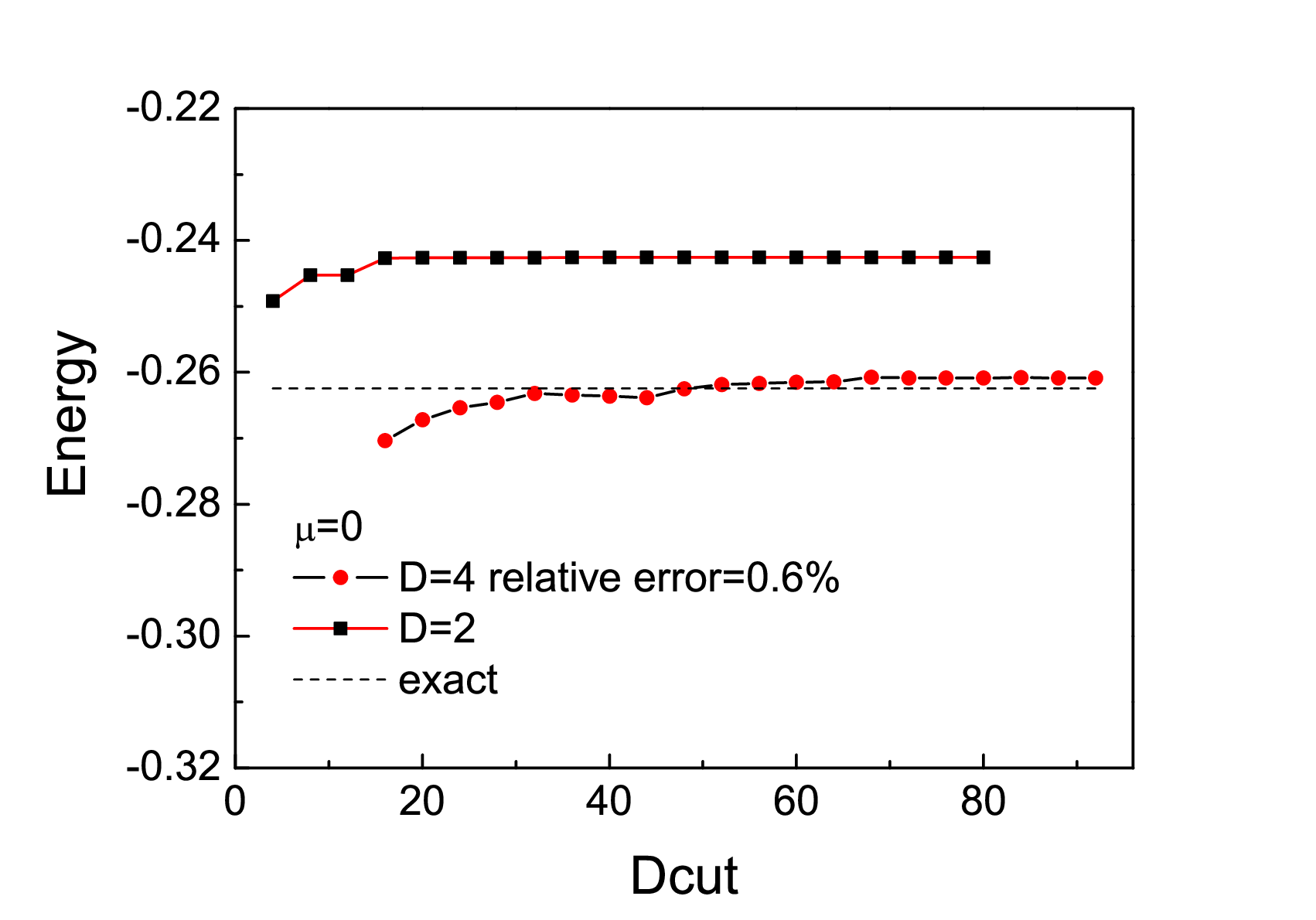}
\end{center}
\caption{(Color online)Ground state energy per link as a function of $D_{cut}$ for fixed
 $\mu=0$ and $\Delta=0.25$. As a bench mark, we also plot the exact energy of the
corresponding free fermion Hamiltonian. \label{freegapless}}
\end{figure}

Although we get good results for the above simple example, it is not
quite surprising since a trivial gapped fermion system only involves
local physics. A much more challenging and interesting example is at
$\mu=0$, in which case the low energy physics of the system is
described by two Dirac cones in the first Brillouin zone(BZ).
Actually, up to a particle-hole transformation on one sublattice,
the above fermion paring Hamiltonian is equivalent to the fermion
hopping Hamiltonian that describes the novel physics in graphene(the
spin less version) where low energy electrons deserve a Dirac like
dispersion. In Fig. \ref{freegapless}, we plot the ground state
energy as a function of $D_{cut}$ for inner dimensions $D=2$ and
$D=4$. In this case, the $D=2$ approach only gives very poor results
comparing to the gapped case. However, when we increase the inner
dimension to $D=4$, we find that the ground state energy agrees
pretty well with the exact one, with a relative error of $0.6\%$.
Another interesting feature is that such a critical system would
require a much larger $D_{cut}$(around $D_{cut}=70$ for $D=4$.) for
the convergence of the GTERG algorithm. Later, we will discuss a
possible improvement for the GTERG algorithm, which allows us to
access much larger inner dimension $D$.

Finally, we would like to make some comments and discussions about
the above results:

(a) Although a gapless fermion system with Dirac cones is critical,
it does not violate the area law because it only contains zero
dimensional Fermi surfaces.

(b) The good agreement in ground state energy does not imply the
good agreement in long range correlations. Indeed, our conjecture is
that for any finite $D$, the variational wave function we derive may
always be associated with a finite correlation length which scales
polynomially in $D$. We note that an interesting critical fPEPS
state with finite inner dimension $D=2$ was proposed in
Ref.\cite{FrankfPEPS}, however, that model is different from our
case since the Dirac cones in that model contain a quadratic
dispersion along some directions in the first BZ.

(c) Although the GTPS variational approach can not describe the
above system with finite inner dimension $D$, the variational
results will still be very useful. From the numerical side, we can
apply both finite $D$ and finite size scalings to estimate the
physical quantities in the infinite $D$ and infinite size limit.
From the analytic side, the finite correlation length corresponds to
a finite gap in the system, which is known as Dirac mass term in the
effective theory\cite{QFT}. In quantum field theory, a controlled
calculation(without singularities in calculating correlation
functions) can be performed by first taking the Dirac mass term to
be finite and then pushing it to the zero limit.

\subsection{An interacting fermion example}
\begin{figure}
\begin{center}
\includegraphics[scale=0.3]{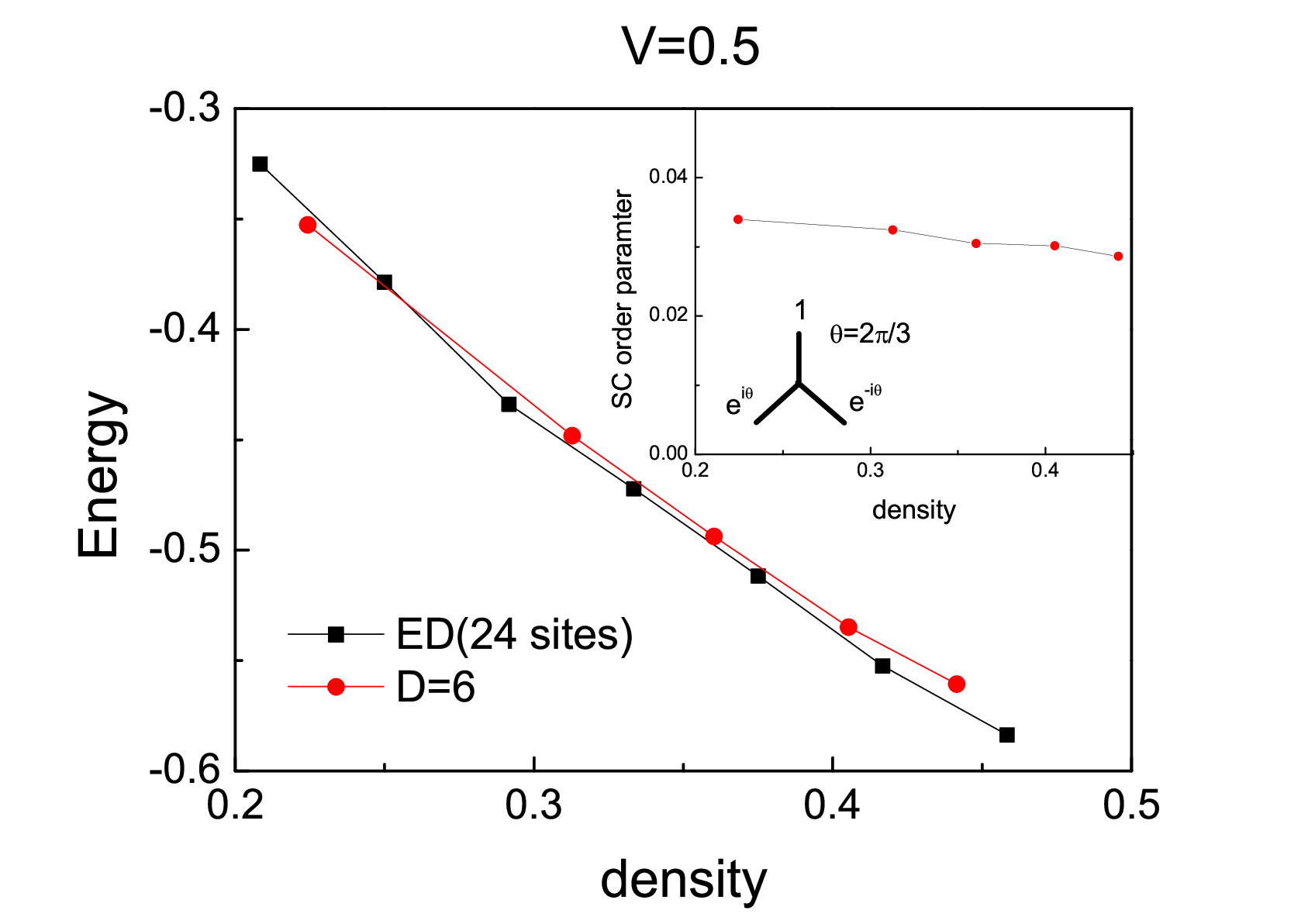}
\end{center}
\caption{A comparison of the ground state energy per link as a function of
electron density for spinless fermion on honeycomb lattice with attractive interactions.}\label{SCenergy}
\end{figure}

To this end, let us test an interacting fermion example with the following Hamiltonian:
\begin{align}
H=-\sum_{\langle ij \rangle}\left(c_i^\dagger
c_j+h.c.\right)-V\sum_{\langle ij \rangle} n_in_j+\mu\sum_i (n_i-n_f) \label{interfermion}
\end{align}
Such a Hamiltonian describes a spinless fermion system on honeycomb lattice with attractive interactions. In Fig. \ref{SCenergy}, we
compare the ground state energy with exact diagonalization(ED) on $24$ sites and find a very good agreement for various
fermion density $n_f$ at $V=0.5$. We use the GTPS ansatz with inner dimension $D=6$. We keep $D_{cut}$ up to $132$ to ensure the convergence of ground state energy. We also compute the superconducting order parameter on nearest neighbor(NN) bond and find a
$p+ip$ pairing symmetry, which is consistent with the general argument that a spinless fermion system with attractive interactions
supports $p+ip$ paring. In the insert of Fig. \ref{SCenergy}, we plot the amplitude of superconducting order parameter for various fermion densities. In Table \ref{SCphase}, we show the phase shift of superconducting order parameters along three primary directions
of honeycomb lattice.

\begin{table}[h]
\begin{tabular}{|c||c|c|c|}
\hline
Doping & $n_f=0.224$ & $n_f=0.313$ & $n_f=0.36$  \\
\hline
$\Delta^{SC}_a/\Delta^{SC}_b$ & (-0.4996,0.8656)   &  (-0.4995,0.8657) & (-0.4995,-0.8656) \\
$\Delta^{SC}_b/\Delta^{SC}_c$ & (-0.5005,0.8660)   &  (-0.5006,0.8659) & (-0.5006,-0.8659) \\
$\Delta^{SC}_c/\Delta^{SC}_a$ & (-0.4999,0.8664)   &  (-0.4999,0.8665) & (-0.4999,-0.8666) \\
\hline
\end{tabular}
\caption{The phase shift of superconducting order parameter along three primary directions on honeycomb lattice.}\label{SCphase}
\end{table}

\section{Possible improvements of the GTERG approach}
\begin{figure}
\begin{center}
\includegraphics[scale=0.45]{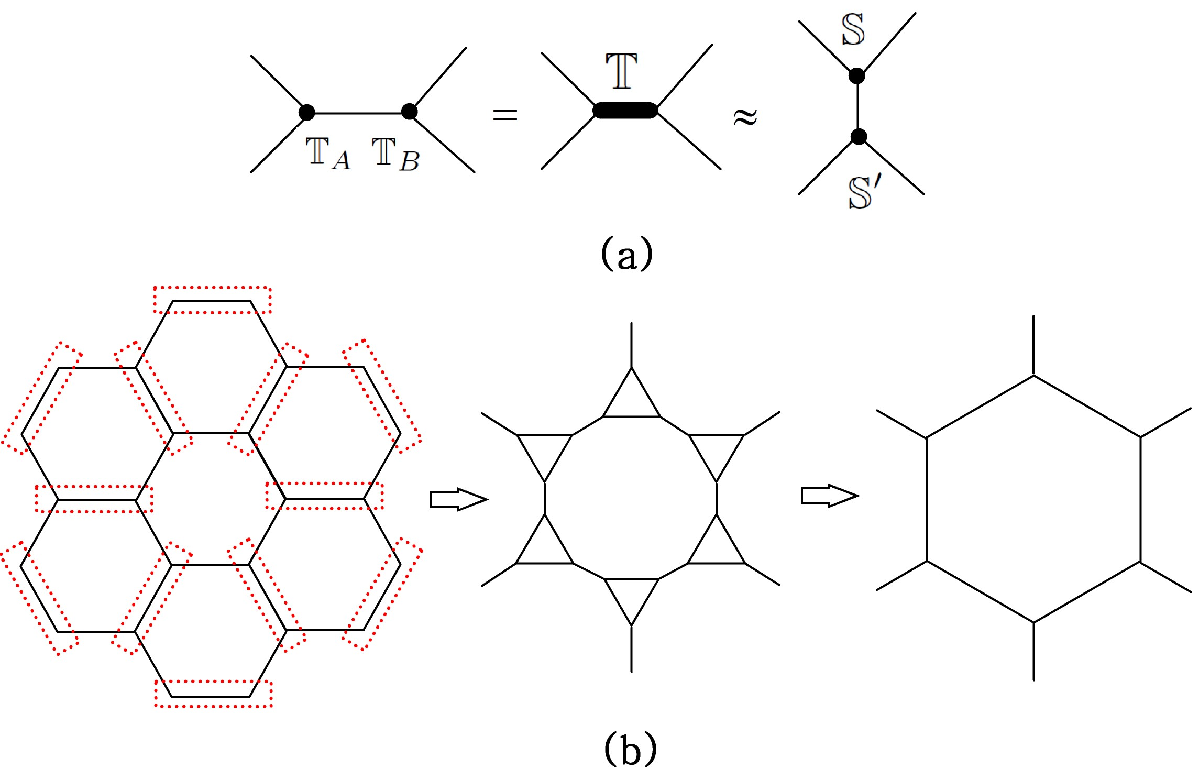}
\end{center}
\caption{(Color online)(a) A graphic representation for
$\mathbb{T}=\mathbb{T}_A\cdot\mathbb{T}_B$ and
$\mathbb{T}\sim\mathbb{S}\cdot\mathbb{S}^\prime$. The symbol $\cdot$
here means sum over the indices for the shared link for two
connected tensors. (b)A schematic plot for the TERG scheme on
honeycomb lattice. The first step is approximate while the second
step is exact.\label{honeycombRG}}
\end{figure}

\begin{figure}
\begin{center}
\includegraphics[scale=0.45]{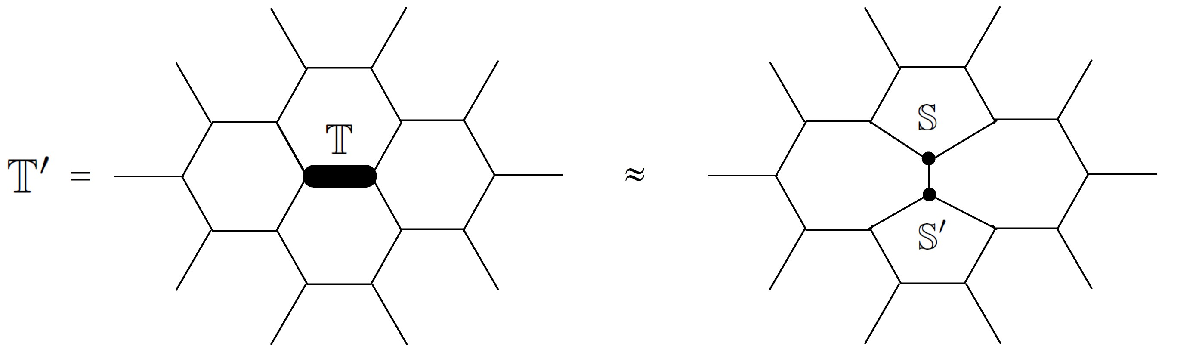}
\end{center}
\caption{A schematic plot for the patch double tensor
$\mathbb{T}^\prime$. The global cost function aims at finding the
best way to minimize the error for the above patch double
tensor.\label{tensortrace}}
\end{figure}

In this section, we will discuss possible improvements for the TERG
algorithm and its Grassmann generalization. Again, let us start from the
usual TPS case. Its Grassmann generalization would be
straightforward by replacing the complex number valued tensors with
those Grassmann valued tensors. We notice that the simple SVD method used
in the TERG algorithm actually implies the following cost function:
\begin{align}
f_{\rm{SVD}}=||\mathbb{T}-\mathbb{S}\cdot\mathbb{S}^\prime||,\label{costSVD}
\end{align}
where the rank four double tensor $\mathbb{T}$ is defined as
$\mathbb{T}_A\cdot\mathbb{T}_B$ and $\cdot$ means summing over
indices for the connected link, as seen in Fig.\ref{honeycombRG}
(a).(Actually it is the inner product of two vectors if we
interpolate $\mathbb{T}_{A(B)}$ as $D^6$ dimensional vectors, where
$D$ is the inner dimension of the TPS.) $||\cdots||$ is the usual
$2$-norm of a vector. (The rank $4$ tensor $\mathbb{T}$ and rank $3$
tensors $\mathbb{S},\mathbb{S}^\prime$ can be viewed as $D^8$ and
$D^6$ dimensional vectors.) Such a cost function only minimizes the
$2$-norm of local error
$\delta\mathbb{T}=\mathbb{T}-\mathbb{S}\cdot\mathbb{S}^\prime$ for a
given cut-off dimension $D_{cut}$(the dimensional of the link shared
by $\mathbb{S}$ and $\mathbb{S}^\prime$) and could not be the
optimal one for minimizing global error. To figure out the optimal
cost function, let us divide the system into large patches, as seen
in Fig.\ref{tensortrace}. If we trace out all the internal indices
inside the patch, we can derive a rank $L$($L$ is the number of
sites on the boundary of the patch) tensor $\mathbb{T}^\prime$ for
the patch and the norm of TPS can be represented as the tensor trace
of the new double tensors $\mathbb{T}^\prime$. When we perform the
TERG algorithm and replace $\mathbb{T}$ by
$\mathbb{S}\cdot\mathbb{S}^\prime$, we aim at making a small error
for the double tensor $\mathbb{T}^\prime$.

Now it is clear why the simple SVD method that minimizes local
errors could not approximate the patch double tensor
$\mathbb{T}^\prime$ in an optimal way. Let us rewrite
$\mathbb{T}^\prime$ as
$\mathbb{T}^\prime=\mathbb{E}\cdot\mathbb{T}$.(Notice that here we regard
the rank $L$ double tensor $\mathbb{T}^\prime$ as a $D^{2L}$
dimensional vector, the rank $4$ double tensor $\mathbb{T}$ as a
$D^8$ dimensional vector and the rank $L+4$ double tensor
$\mathbb{E}$ as a $D^{2L}$ by $D^8$ matrix.) The double tensor
$\mathbb{E}$ is called environment tensor. The cost function
which provides the best approximation for $\mathbb{T}^\prime$ can be
represented as:
\begin{align}
f=||\mathbb{E}\cdot\left(\mathbb{T}-\mathbb{S}\cdot\mathbb{S}^\prime\right)||
\end{align}
Again, $\cdot$ means the vector inner product and $||\cdots||$ means
the usual $2$-norm. Comparing to the simple SVD method, the
environment tensor gives a complex weight for each component of the
local error $\delta\mathbb{T}$. Notice that the cost function proposed
here is very different from the one in Ref.\cite{XiangTRG2}, which
aims at minimizing
${\rm{Tr}}\left(\mathbb{E}\cdot\mathbb{T}\right)$.

\begin{figure}
\begin{center}
\includegraphics[scale=0.35]{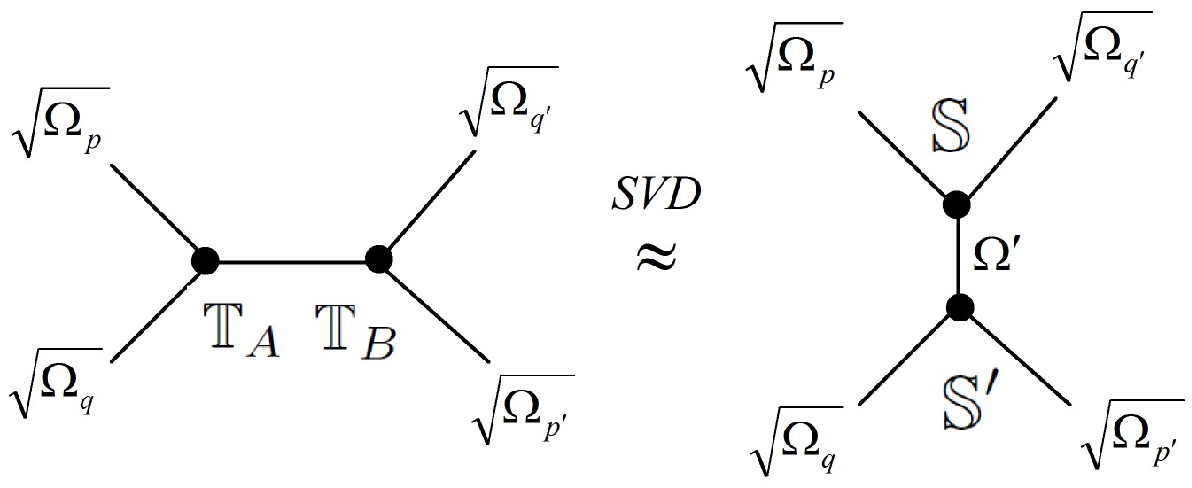}
\end{center}
\caption{A graphic representation for the first step of the weighted
TERG algorithm.\label{weightSVD}}
\end{figure}

Although the environment tensor $\mathbb{E}$ is conceptually useful,
it is impossible to compute this rank $L+4$ tensor when the patch
size becomes very large. Nevertheless, we can derive a simplified
environment through the simplified time evolution algorithm. The key
step is also based on the conjecture that the rank $8$ tensor
$\mathbb{E}^\dagger \cdot \mathbb{E}$ can be factorized into a
product form:
\begin{align}
\left(\mathbb{E}^\dagger \cdot \mathbb{E}\right)_{pqp^\prime
q^\prime ;\bar p \bar q \bar p^\prime \bar q^\prime
}=\Omega_{p}^y\Omega_{q}^z\Omega_{p^\prime}^y\Omega_{q^\prime}^z
\delta_{p \bar p}\delta_{q \bar q}\delta_{p^\prime \bar
p^\prime}\delta_{p^\prime \bar p^\prime},\label{environment}
\end{align}
where $p=(b,\bar b)$,$q=(c,\bar c)$ are the double indices. For any
converged $T_A$ and $T_B$ from the simplified time evolution
algorithm with converged weight vectors ${\Lambda^*}^\alpha$, the
weight vectors $\Omega^\alpha$ for double tensors $\mathbb{T}$ can
be determined as:
\begin{align}
\Omega_{p}^y={\Lambda^*}_{b}^y{\Lambda^*}_{\bar b}^y,
\Omega_{q}^z={\Lambda^*}_{c}^z{\Lambda^*}_{\bar c}^z.
\end{align}
The above conjecture for the environment is reasonable when the
environment of $T^m$ can be approximated as a product form since
$\mathbb{T}=\sum_m {\bar T^m}\otimes T^m$

Similar to the simplified time evolution algorithm, solving the
optimal cost function in this case can be implemented by the SVD
decomposition for matrix $\mathbb{M}$(see Fig. \ref{weightSVD}):
\begin{align}
\mathbb{M}_{q^\prime p;qp^\prime }=&\mathbb{T}_{pqp^\prime
q^\prime}\sqrt{\Omega_p}\sqrt{\Omega_q}\sqrt{\Omega_{p^\prime}}\sqrt{\Omega_{q^\prime}}\nonumber\\
\simeq & \sum_{r=1}^{D_{cut}} U_{q^\prime p;r}V_{qp^\prime
;r}\Omega_r^\prime
\end{align}
Then the rank $3$ double tensors $\mathbb{S}$ and
$\mathbb{S}^\prime$ can be determined as:
\begin{align}
\mathbb{S}_{rq^\prime p}&=\frac{\sqrt{\Omega_r^\prime}}{\sqrt{\Omega_{q^\prime}}\sqrt{\Omega_p}}U_{q^\prime p;r}\nonumber\\
\mathbb{S}_{r^\prime qp^\prime}^\prime
&=\frac{\sqrt{\Omega_{r^\prime}^\prime}}{\sqrt{\Omega_{q}}\sqrt{\Omega_{p^\prime}}}V_{qp^\prime;r^\prime}
\end{align}
Finally, the weight vector $\Omega^x$ for $x-link$ can be updated as
$\Omega^x=\Omega^\prime$.

In general cases, the assumption Eq.(\ref{environment}) could not be
true, however, as long as $\mathbb{E}_{\lambda;pqp^\prime
q^\prime}/\sqrt{\Omega_p}\sqrt{\Omega_q}\sqrt{\Omega_{p^\prime}}\sqrt{\Omega_{q^\prime}}$
has a much more uniform distribution(up to proper normalization):
\begin{align}
\left|\frac{\mathbb{E}_{\lambda;pqp^\prime
q^\prime}}{\sqrt{\Omega_p}\sqrt{\Omega_q}\sqrt{\Omega_{p^\prime}}\sqrt{\Omega_{q^\prime}}}\right|\sim
1,
\end{align}
the above weighted TERG(wTERG) algorithm can still improve the
accuracy for fixed $D_{cut}$ with the same cost. This is because the
SVD method is the best truncation method if the environment has a
random but uniform distribution.

By replacing the complex valued double tensors with Grassmann
variable valued double tensors, all the above discussions will be
valid for GTPS. However, the definition of the inner product $\cdot$
and the corresponding 2-norm $||\cdots||$ should also be generalized
into their Grassmann version, which evolves the integration over
Grassmann variables for the connected links with respect to the
standard Grassmann metric. For example, the cost function of the
GSVD method discussed in Ref.\cite{GuGTPS} can be formally written
as:
\begin{align}
f_{\rm{GSVD}}=\delta\mathbb{T}^f\cdot{\delta\mathbb{T}^f}^\dagger=
||\delta\mathbb{T}^f
||=||\mathbb{T}^f-\mathbb{S}^f\cdot{\mathbb{S}^\prime}^f||,\label{costGSVD}
\end{align}
To explain the meaning of the above expression more explicitly, we
consider a simple case that $\mathbb{T}^f$ only contains one species
of Grassmann variables. We can express $\mathbb{T}^f$ as:
\begin{align}
\mathbb{T}^f_{pqp^\prime q^\prime}=\mathbb{T}_{pqp^\prime
q^\prime}\left(\theta_\beta\right)^{N_f(p)}\left(\theta_\gamma\right)^{N_f(q)}
\left(\theta_{\beta^\prime}\right)^{N_f(p^\prime)}\left(\theta_{\gamma^\prime}\right)^{N_f(q^\prime)},
\end{align}
where $\mathbb{T}_{pqp^\prime q^\prime}$ is the complex coefficient
of the Grassmann number valued double tensor
$\mathbb{T}^f_{pqp^\prime q^\prime}$ and $N_f(p)=\frac{P_f(p)+1}{2}$
is determined by the fermion parity of the inner index $p$. We
notice that $\mathbb{T}_{pqp^\prime q^\prime}=\sum_r \mathbb{T}_{A;rpq
}\mathbb{T}_{B;rp^\prime q^\prime}$ if we express the Grassmann
number valued double tensors $\mathbb{T}_{A(B);rpq }^f$ on
sublattice $A(B)$ as:
\begin{align}
\mathbb{T}_{A(B);rpq }^f=\mathbb{T}_{A(B);rpq
}\left(\theta_\alpha\right)^{N_f(r)}\left(\theta_\beta\right)^{N_f(p)}\left(\theta_{\gamma}\right)^{N_f(q)}
\end{align}
Similarly, we can express $\mathbb{S}^f$ and $\mathbb{S^\prime}^f$
as:
\begin{align}
\mathbb{S}^f_{rq^\prime p}&=\mathbb{S}_{rq^\prime
p}\left(\theta_\alpha\right)^{N_f(r)}\left(\theta_{\gamma^\prime}\right)^{N_f(q^\prime)}\left(\theta_\beta\right)^{N_f(p)}
\nonumber\\
\mathbb{S^\prime}^f_{ r^\prime q p^\prime
}&=\mathbb{S^\prime}_{r^\prime q p^\prime}
\left(\theta_{\alpha^\prime}\right)^{N_f(r^\prime)}
\left(\theta_\gamma\right)^{N_f(q)}\left(\theta_{\beta^\prime}\right)^{N_f(p^\prime)}
\label{Gdecompose}
\end{align}
Again, $\mathbb{S}_{rq^\prime p}$ and $\mathbb{S^\prime}_{r^\prime q
p^\prime}$ are the complex valued coefficients. Recall the
definition of the standard Grassmann metric
$\mathbf{g}_{rr^\prime}$:
\begin{align}
\mathbf{g}_{rr^\prime}=\delta_{rr^\prime}
\left(\dd\theta_\alpha\right)^{N_f(r)}\left(\dd\theta_{\alpha^\prime}\right)^{N_f(r^\prime)},
\end{align}
the inner product $\mathbb{S}^f\cdot\mathbb{S^\prime}^f$ explicitly
means:
\begin{widetext}
\begin{align}
\left(\mathbb{S}^f \cdot \mathbb{S^\prime}^f \right)_{pqp^\prime
q^\prime} &=\sum_{rr^\prime} \int \mathbf{g}_{rr^\prime}
\mathbb{S}_{rq^\prime p}\mathbb{S^\prime}_{ r^\prime q
p^\prime}\left(\theta_\alpha\right)^{N_f(r)}\left(\theta_{\gamma^\prime}\right)^{N_f(q^\prime)}\left(\theta_\beta\right)^{N_f(p)}
\left(\theta_{\alpha^\prime}\right)^{N_f(r^\prime)}
\left(\theta_\gamma\right)^{N_f(q)}\left(\theta_{\beta^\prime}\right)^{N_f(p^\prime)}
\nonumber\\
&= \sum_{r}  \mathbb{S}_{rq^\prime p}\mathbb{S^\prime}_{ r^\prime q
p^\prime}
\left(\theta_{\gamma^\prime}\right)^{N_f(q^\prime)}\left(\theta_\beta\right)^{N_f(p)}
\left(\theta_\gamma\right)^{N_f(q)}\left(\theta_{\beta^\prime}\right)^{N_f(p^\prime)}.
\end{align}
\end{widetext}
Thus, we have:
\begin{align}
\delta\mathbb{T}^f_{pqp^\prime
q^\prime}=\delta\mathbb{T}_{pqp^\prime
q^\prime}\left(\theta_{\gamma^\prime}\right)^{N_f(q^\prime)}\left(\theta_\beta\right)^{N_f(p)}
\left(\theta_\gamma\right)^{N_f(q)}\left(\theta_{\beta^\prime}\right)^{N_f(p^\prime)},
\end{align}
with
\begin{align}
\delta\mathbb{T}_{pqp^\prime q^\prime}&=\mathbb{T}_{pqp^\prime
q^\prime}(-)^{N_f(p^\prime)}-\sum_{r} \mathbb{S}_{rq^\prime
p}\mathbb{S^\prime}_{ r q
p^\prime}\nonumber\\
&=\mathbb{T}_{pqp^\prime q^\prime}^\prime-\sum_{r}
\mathbb{S}_{rq^\prime p}\mathbb{S^\prime}_{ r q p^\prime}
\end{align}
Now it is clear that up to a sign twist, the GSVD cost function is
equivalent to the cost function of its complex coefficient tensors:
\begin{align}
f_{GSVD}=||\delta\mathbb{T}^f||=||\delta\mathbb{T}||=||\mathbb{T}^\prime-\mathbb{S}\cdot{\mathbb{S}^\prime}||
\end{align}
Explicitly the same as in the TERG case, the best approximation for
a given $D_{cut}$ is nothing but the SVD decomposition for the
coefficient tensor $\mathbb{T}^\prime$.(If we view
$\mathbb{T}_{pqp^\prime q^\prime}^\prime$ as a matrix
$\mathbb{M}_{q^\prime p;qp^\prime}=\mathbb{T}_{pqp^\prime
q^\prime}^\prime\simeq \sum_{r=1}^{D_{cut}}\mathbb{S}_{rp^\prime
q}\mathbb{S^\prime}_{ r q p^\prime}$.) On the other hand, as already
having been discussed in Ref.\cite{GuGTPS}, the constraint Eq.
(\ref{constraint}) of GTPS implies that their double tensors contain
even number of Grassmann numbers. As a result, $\mathbb{M}$ is block
diagonalized and the index $r$ will have a definite parity
$P^f(r)=P^f(p)P^f(q)=P^f(p^\prime)P^f(q^\prime)$. We notice that the
novel sign factor $(-)^{N_f(p^\prime)}$ here arises from the
anti-commuting nature of the Grassmann variables and will encode the
fermion statistics. The second RG step remains the same as in GTERG
and a similar novel sign factor will also emerge there.

All the above discussion will still be correct if the inner index of
the double tensor $\mathbb{T}^f$ contains multiple species of
Grassmann variables. Indeed, starting from the standard GTPS, the
double tensor $\mathbb{T}^f$ in the first RG step will contain two
species of Grassmann variables. This is because
$\mathbb{T}^f_{A(B)}=\sum_m \mathbf{\bar T}_{A(B)}^m \otimes
\mathbf{T}_{A(B)}^m $ and $\mathbb{T}^f=\mathbb{T}^f_A\cdot
\mathbb{T}^f_B$. In the second and latter RG steps,
$\mathbb{T}^f_{A(B)}$ will only contain one species of Grassmann
variables.

The discussion for the environment effect will be in a similar
way. Especially, if we use some weighting factors to approximately represent
the environment effect, the coefficient tensors
$\mathbb{S},\mathbb{S}^\prime$ will take the same form as in the usual
TPS case. However, an important sign factor should be included when
we define the matrix $\mathbb{M}$. Again,the environment weight for
the first step is determined by the time evolution algorithm of
GTPS. Same as in the bosonic case, the singular value obtained from
the GSVD would perform as the environment weight for the next RG step.

\begin{figure}
\begin{center}
\includegraphics[scale=0.33]{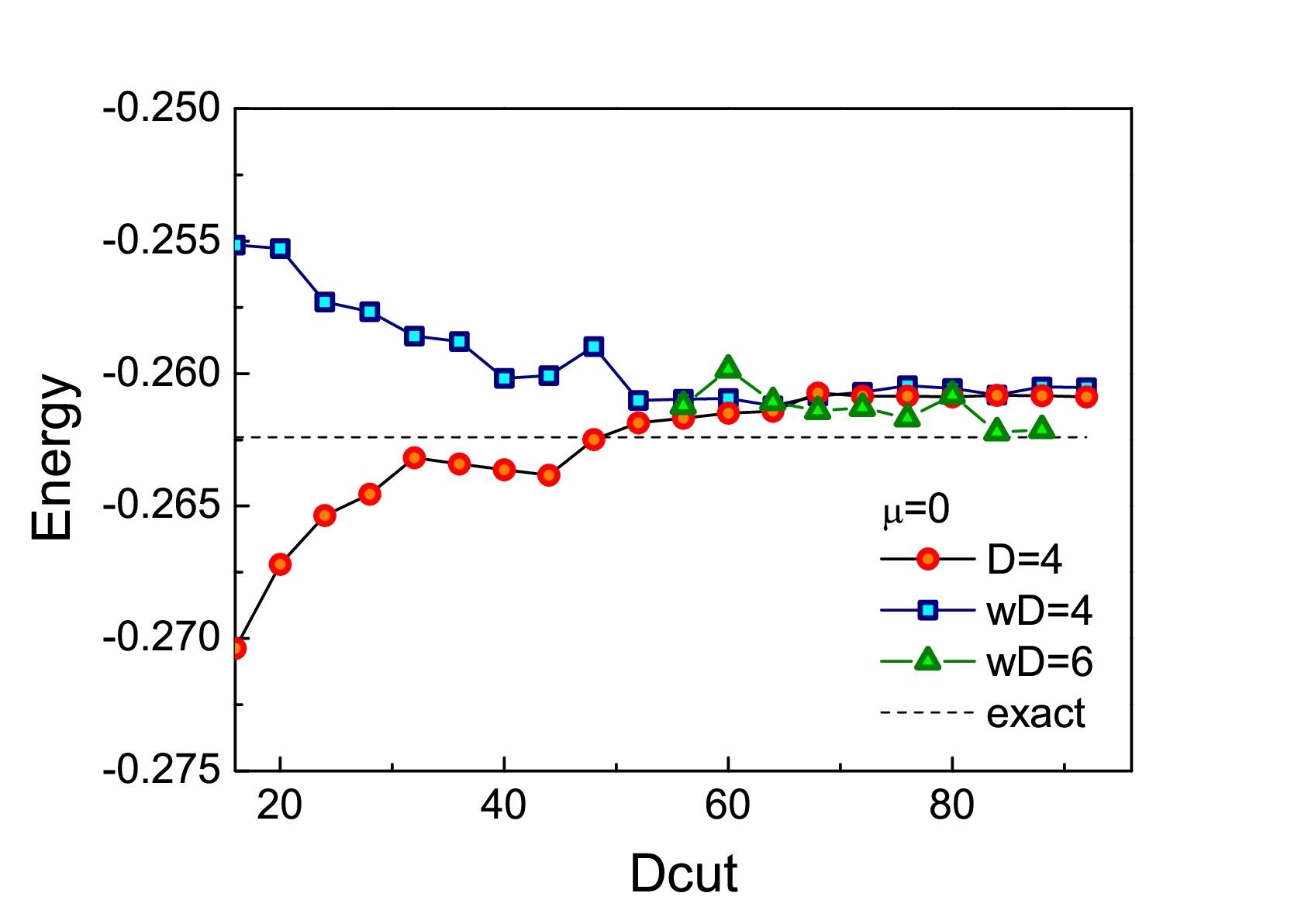}
\end{center}
\caption{A comparison of the ground state energy per link as a function of
$D_{cut}$ at $\mu=0$ and $\Delta=0.25$ . By introducing the environment weight in the
GTERG method, an important improvement is that the ground state
energy is always above the exact value, which is very important for
variational approach}.\label{gaplessweight}
\end{figure}

In Fig. \ref{gaplessweight}, we implement the above algorithm to the
free fermion Hamiltonian Eq.(\ref{freefermion}) at critical
point($\mu=0$). We see an important improvement that the ground
state energy decreases when increasing $D_{cut}$ and is strictly
above the exact energy, unlike the simple GTERG approach, which can
overestimate the ground state energy for small $D_{cut}$. However,
for large enough $D_{cut}$, the two approaches converge to the same
values, as expected. We further use the new algorithm to study the
critical model with larger inner dimension $D$. Up to $D=6$, we find
that the ground state energy from the GTPS approach is almost the
same as the exact one(relative error $\sim 0.1\%$).

\section{Summary}
In this paper, we first derive a standard form of GTPS that only
contains one species of Grassmann variables for each inner index and
significantly simplifies the representations in our numerical
calculations. Based on the fermion coherent state representation, we
further generalize the imaginary time evolution algorithms into
fermion systems. We study a simple free fermion example on honeycomb
lattice, including both off-critical and critical cases to test our
new algorithms. Finally, we discuss the importance of the
environment effect of the TERG/GTERG method and present a simple
improvement by introducing proper environment weights.

Although the simple time evolution algorithm discussed here is not
generic enough, it has already allowed us to study many interesting
and important models, such as the Hubbard/$t-J$ model, whose ground
state is believed to be a superconductor. The evidence for the
existence of superconductivity in these models based on the GTPS
algorithm will be discussed and bench marked with other methods
elsewhere\cite{honeycombtJ}. Of course, the generic algorithm is
also very important and desired, especially for those systems with
topological order. Actually, the general discussions in Section II
have already made some progress along this direction, but not
efficient and stable enough at this stage.

On the other hand, further improving the efficiency of contracting
(Grassmanna) tensor net is also very important. Although the
GTERG/TERG algorithm provides us promising results in many cases, it
is still not efficient enough since the algorithm is not easy to be
parallelized. Recently, a novel idea of combining the concept of
renormalization and Monte Carlo(MC)\cite{LingMCTRG} has made great
success for boson/spin systems, it would be very natural to
generalize it into fermion/electron systems based on the Grassmann
variable representations.

We would like to thank F.Verstrate, J.I.Cirac and X.G. Wen for very
helpful discussions. We especially thank D.N. Sheng for providing the ED data.
 This work is supported in part by NSF Grant
No. NSFPHY05-51164

\end{document}